\documentclass[aps,pre,groupedaddress,twocolumn,floatfix]{revtex4-2}
\usepackage{xcolor}

\newcommand{\beq}{\begin{equation}}
\newcommand{\eeq}{\end{equation}}
\def\NOTE#1{\textcolor{black}{#1}}
\def\NOTEE#1{\textcolor{black}{#1}}

\usepackage{amsmath,amssymb}
\usepackage{graphicx}
\usepackage{lipsum}

\begin{document}

%\preprint{}

\title{Intermittency of three-dimensional perturbations in a point-vortex model}


\author{Adrian van Kan} \email[]{avankan@ens.fr}
\author{Alexandros Alexakis} \email[]{alexakis@phys.ens.fr} \author{Marc-Etienne Brachet} \email[]{marc.brachet@gmail.com}
%\altaffiliation{D\'{e}partement de G\'{e}osciences, \'{E}cole Normale Sup\'{e}rieure, PSL Research University, Paris, France}
\affiliation{Laboratoire de Physique de l’Ecole normale supérieure, ENS, Université PSL, CNRS, Sorbonne Université, Université de Paris, F-75005 Paris, France}


\date{\today}

\begin{abstract}
Three-dimensional (3-D) instabilities on a (potentially turbulent) two-dimensional (2-D) flow are still incompletely understood, despite recent progress. Here, \NOTEE{based on known physical properties of such 3-D instabilities}, we propose a simple, energy-conserving model \NOTEE{describing this situation. It consists of a} regularized 2-D point-vortex flow coupled to localized 3-D perturbations (``{\it ergophages}"), such that  {\it ergophages} can gain energy by altering vortex-vortex distances through an induced divergent velocity field, thus decreasing point-vortex energy.
We investigate the model in three distinct stages of evolution:  
%linear, passive nonlinear and fully nonlinear evolution of 3-D instabilities in the model. 
%
(i) The {\it linear regime}, where the amplitude of the {\it ergophages} grows or decays exponentially on average, with an instantaneous growth rate that fluctuates randomly in time. The instantaneous growth rate has a small auto-correlation time, and a probability distribution featuring a power-law tail with exponent between $-2$ and $-5/3$ (up to a cut-off) depending on \NOTEE{the point-vortex base flow}. %background state of the point-vortex flow. 
%These power-laws can be predicted directly from the model equations.
Consequently, the logarithm of the {\it ergophage} amplitude performs a %free
L\'evy flight. 
%This novel application of L\'evy flights suggests an explanation for the jump-like behavior recently observed in direct numerical simulations (DNS) of 3-D instabilities on turbulent 2-D flow in \cite{seshasayanan2020onset}. 
%In particular, the presence of L\'evy flights suggests a significant difficulty for direct numerical simulations (DNS) to obtain correct statistics of 3-D perturbations, since the power-law tails correspond to diverging mean and higher moments in the absence of a cut-off. 
(ii) The {\it passive-nonlinear} regime of the model, where the 2-D flow evolves independently of the {\it ergophage} amplitudes, which saturate by non-linear self-interactions without affecting the 2-D flow. In this regime the system exhibits a new type of on-off intermittency that we name \textit{L\'evy on-off intermittency},
which we define and study in a companion paper [van Kan et al. 2021]. We compute the bifurcation diagram for the mean and variance of the perturbation amplitude, as well as the probability density of the perturbation amplitude.
%both from the model and from integrating a corresponding Langevin equation with truncated L\'evy noise. 
%We find that as a consequence of the truncation, the results agree with the predictions for on-off intermitteny with Gaussian noise. 
(iii) Finally, we characterize the {\it fully  nonlinear regime}, where {\it ergophages} feed back on the 2-D flow, and study how the vortex temperature is altered by the interaction with ergophages. It is shown that when the amplitude of the {\it ergophages} is sufficiently large, the \NOTEE{condensate is disrupted and the} 2-D flow saturates to a zero-temperature state.
Given the limitations of existing theories, our model provides a new perspective on 3-D instabilities growing on 2-D flows, which will be useful in analysing and understanding the much more complex results of DNS and potentially guide further theoretical developments.
\end{abstract}

%\keywords{}

\maketitle

% \begin{figure}
% \end{figure}

% Surround figure environment with turnpage environment for landscape
% figure
% \begin{figure}
% \end{figure}

%\begin{acknowledgments}
%\end{acknowledgments}

\section{Introduction}
\label{sec:intro}
Point-vortex flow is a simple (but singular, \NOTEE{i.e. weak}) solution of the two-dimensional (2-D) Euler equation describing inviscid fluid flow, in which $N$ strongly localized vortices advect each another chaotically by their induced velocity fields \cite{helmholtz1867integrals,kirchhoff1876vorlesungen,saffman1986difficulties,greengard1988singular,goodman1990convergence}. They admit a famous equilibrium statistical mechanics description due to Onsager \cite{onsager1949statistical, eyink2006onsager}, who showed that states with negative temperatures exist in the system, where same-signed point vortices cluster to form two strong counter-rotating vortices. Indeed, 2-D turbulent flow features isolated vortices which aggregate and merge over time in a process called \textit{inverse energy cascade}, forming a large-scale \textit{condensate}, where most of the energy is concentrated in the largest-scale mode \cite{kraichnan1980two,tabeling2002two,bofetta2012twodimensional}. This is in contrast with three-dimensional (3-D) turbulence, where energy is transferred from large to small scales \cite{frisch1995turbulence}. Inverse cascades and associated condensation phenomena are also found in quasi-2-D flows, such as turbulence in thin layers \cite{celani2010turbulence,xia2011upscale,benavides2017critical,van2019condensates,musacchio2019condensate} and rapidly rotating turbulence \cite{smith1996crossover,deusebio2014dimensional}, which feature 3-D components, but are predominantly 2-D. A review of such flows is given in \cite{alexakis2018cascades}. 

Point-vortex models have found numerous applications in simplified descriptions of turbulent fluid flows. An early successful simulation of the inverse cascade in 2-D turbulence indeed relied on the point-vortex-based vortex-in-cell approximation, \cite{siggia1981point}. In the 1990s, there was a significant activity devoted to vortex gas modelling of (particularly decaying) 2-D turbulence \cite{carnevale1991evolution, benzi1992simple,weiss1993temporal,trizac1998coalescence,weiss1999punctuated}, where merging rules for point vortices were prescribed, yielding 2-D turbulence-like behavior at reduced numerical cost. Point-vortex models have also been used to investigate stirring by chaotic advection \cite{aref1984stirring}, as well as Lagrangian intermittency, pair dispersion and transport in turbulence \cite{rast2009point,rast2011pair,rast2016turbulent}. Recently, vortex gas scaling arguments were leveraged to find a highly accurate local closure in baroclinic turbulence \cite{gallet2020vortex}. Other physical problems which have been fruitfully treated by point-vortex models include the stability of vortex streets and vortex sheets \cite{lamb1945hydrodynamics,aref1981evolution,krasny1986study,krasny1986desingularization}, quantum turbulence \cite{nowak2012nonthermal,reeves2013inverse,billam2014onsager,griffin2020magnus}, plasma dynamics \cite{joyce1973negative} and stellar dynamics \cite{chavanis1996statistical}.

For flows in thin layers, rotating flows and flows under the action of an external magnetic field, it has been proven using upper bound theory \cite{gallet2015exact,gallet2015exact2} that a non-dimensional threshold exists in terms of the layer depth and fluid viscosity (as well as the rotation rate and or the external magnetic field, if present), where the flow undergoes exact bi-dimensionalization (for periodic or stress-free boundary conditions). Beyond this point, 3-D perturbations away from a 2-D flow decay due to the action of viscous damping. This has profound consequences for turbulent flows since, as mentioned, the phenomenology of 2-D turbulence differs strongly from the 3-D case due to additional conserved quantities in the 2-D case \cite{frisch1995turbulence,bofetta2012twodimensional}. Therefore, it is important to understand quasi-2-D flows close to the onset of three-dimensionality. The bounding theory only establishes the existence of a threshold, but since it is built on rather conservative estimates, it cannot capture the physics occurring near the threshold. Very recently, in an extensive numerical study \cite{seshasayanan2020onset}, Seshasayanan and Gallet investigated the linear stability of 3-D perturbations on a 2-D turbulent condensate background flow at the onset of three-dimensionality. The authors showed that when instability is present, the time evolution of the energy of linear 3-D modes involves phases of jump-like exponential growth occurring randomly in time, inter-spaced by plateau-like phases where growth is absent. %The authors further show that the probability density function (PDF) of the growth rate of the (domain mean logarithmic) amplitude of the 3-D modes shows power law tails.
Here, in the spirit of the wide range of applications of point vortices described above, we formulate and analyze a point-vortex model of localized 3-D perturbations in quasi-2-D turbulence, whose dynamics are qualitatively similar to the exponential growth and decay evolution found in \cite{seshasayanan2020onset}. 
%\NOTE{We proceed to study the model in three steps: the passive linear, passive nonlinear and fully nonlinear regimes. In the passive linear regime, the perturbation amplitude is small, such that it evolves according to a linear equation, and the 3-D ergophages are passive, such that the 2-D flow evolves independently of 3-D perturbations. In the passive nonlinear regime, the ergophages remain passive, but the perturbation amplitude saturates nonlinearly. In the fully nonlinear regime, the 2-D flow is altered by 3-D perturbations in an energy-conserving manner. We study the statistics of the fluctuating growth rates of 3-D instabilities in the linear regime, as well as the nonlinear saturation of these instabilities in the nonlinear regimes. As will be shown below, the amplitude of 3-D instabilities in the model turns out to perform a L\'evy flight in a quartic potential, which shows intermittent behaviour close to a threshold. This article is accompanied by a companion paper containing a theoretical analysis of this situation, which we baptise \textit{extreme on-off intermittency}, \cite{vankan2020extreme}.}

The remainder of this article is structured as follows. In section \ref{sec:bg_statmech}, \NOTE{we provide a brief introduction to the concept of point-vortex temperature}, in section \ref{sec:model}, we formulate the model to be studied. In section \ref{sec:method}, we describe the method of our investigation. Then, in section \ref{sec:simres} we present the results of our numerical simulations and finally in section \ref{sec:conclusions} we discuss the implications of our results and remaining open questions. 

%%%%%%%%%%%%%%%%%%%%%%%%%%%%%%%%%%%%%%%%%%%%%%%%%%%%%%%%%%%%%%
\section{Background: Temperature of point-vortex states}   %%%
\begin{figure}
    \centering
    \includegraphics[width=8.6cm]{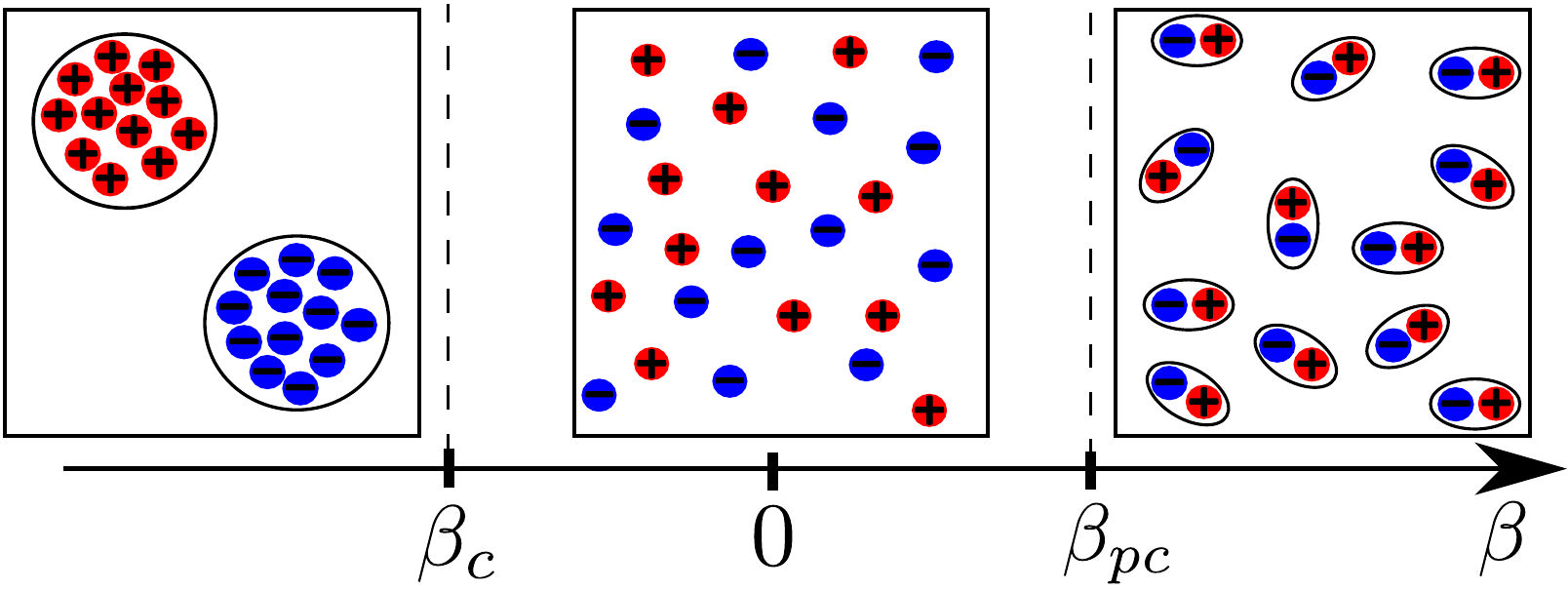}
    \caption{Overview of point-vortex states at negative, zero and positive inverse temperatures $\beta$. Clustering occurs for $\beta<\beta_c<0$, a homogeneous state is found at $\beta=0$, and pair condensation occurs for $\beta>\beta_{pc}$.}
    \label{fig:overview_vortices}
\end{figure}

\begin{figure}[h]
    \centering
     \includegraphics[width=9.0cm]{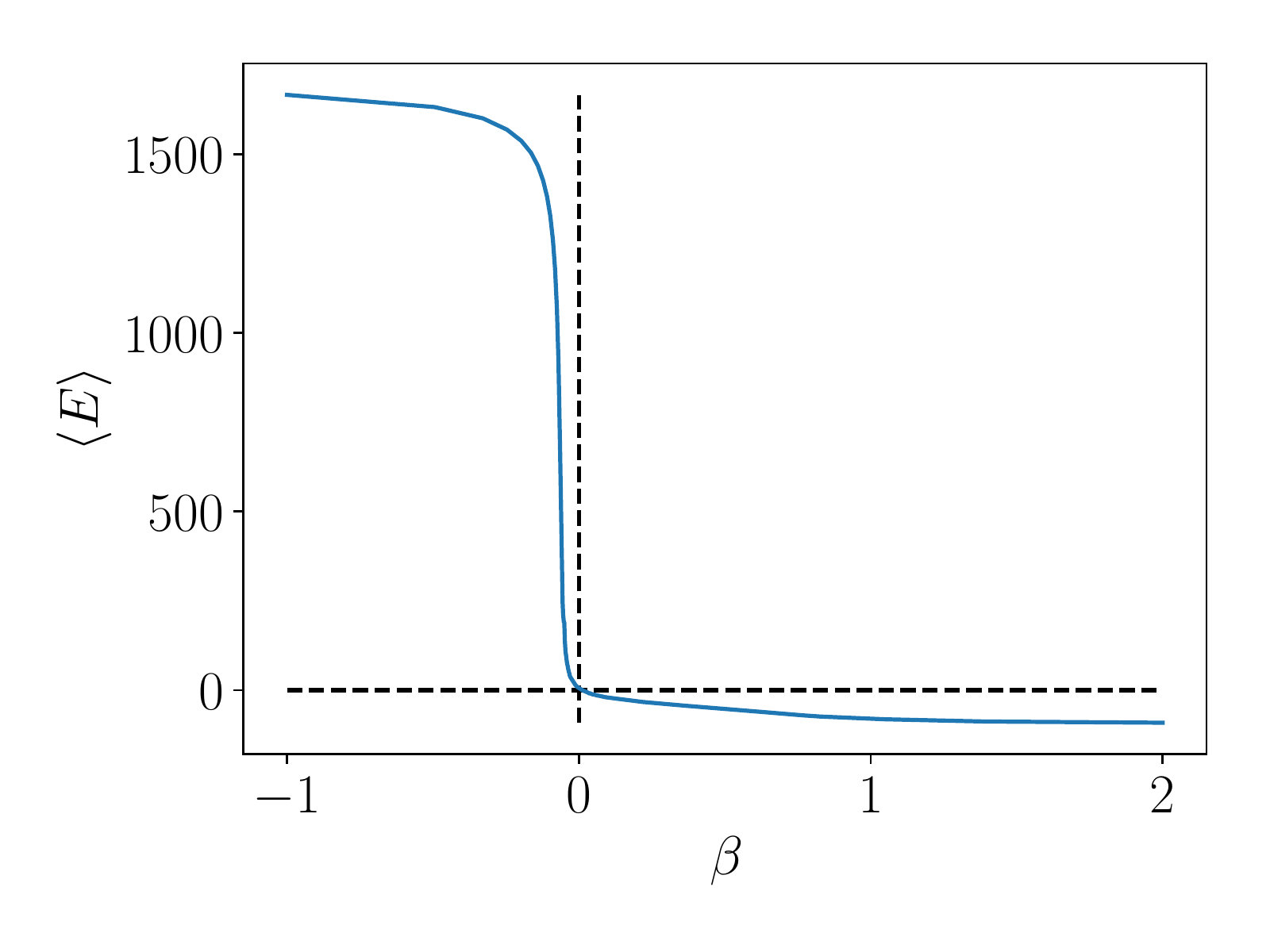} 
    \caption{Mean point-vortex energy  $\langle E \rangle$ of $N_v=32$ vortices versus $\beta$, computed using the method described in appendix \ref{sec:appA} in the periodic domain $[0,2\pi]\times[0,2\pi]$ (with a truncation at distances smaller than $\epsilon=0.1$, cf. appendix \ref{sec:appA}). This curve allows a translation from vortex energies at steady state to corresponding temperatures.}
    \label{fig:E_vs_beta}
\end{figure}

\label{sec:bg_statmech}                                    %%%
%%%%%%%%%%%%%%%%%%%%%%%%%%%%%%%%%%%%%%%%%%%%%%%%%%%%%%%%%%%%%%
We briefly summarize the concept of the temperature of point-vortex flow, which was introduced in 1949 by Onsager \cite{onsager1949statistical}. The energy of a set of point vortices is given by the Hamiltonian $H$, which only depends on the vortex positions $(x,y)$. These positions are the conjugate variables of the point-vortex Hamiltonian. In bounded domains, the total phase space volume is therefore finite. We denote by $\Omega(E)$ the phase space volume occupied by states whose energies $H$ lie in the interval [$E, E+dE$]. Then the thermodynamic entropy is $k_B\ln (\Omega(E)/\Omega_0)$, where $k_B$ is the Boltzmann constant and $\Omega_0$ is a reference volume required for dimensional reasons. In the extreme situation where vortex dipoles (vortex-antivortex pairs) collapse, which corresponds to negative energies $E<0$, the available phase space volume is vanishingly small, $\Omega(E)\stackrel{E\to-\infty} \longrightarrow 0$. The opposite limit of large positive energies occurs when like-sign vortices concentrate at a point, in which case also $\Omega(E)\stackrel{E\to \infty}{\longrightarrow}0$. Since the total volume is non-zero, the non-negative function $\Omega(E)$ must reach a maximum at an intermediate energy $-\infty<E_m<\infty$. The associated microcanonical inverse temperature,
\begin{equation}
    \beta(E) \equiv\frac{\partial \ln(\Omega(E))}{\partial E} \label{eq:def_beta}
\end{equation}
is thus positive for $E<E_m$, but vanishes at $E=E_m$  and is negative for $E>E_m$. Negative-temperature states can generally arise in \NOTE{both classical and quantum systems with a finite number of degrees of freedom whose state space is bounded}, such as localized spin systems \cite{purcell1951nuclear,oja1997nuclear,medley2011spin}. In the point-vortex system, high-energy states at negative temperatures, corresponding to condensates featuring same-sign vortex clusters, have been extensively studied since Onsager's initial contribution \cite{onsager1949statistical,yatsuyanagi2005dynamics,eyink2006onsager,yu2016theory}. In particular, there is a negative clustering temperature $\beta_c$, which marks the onset of same-sign vortex clustering. Similarly, there is a positive pair condensation temperature $\beta_{pc}$, at which opposite-sign vortices form dipole pairs which propagate through the domain, see \cite{cornu1987two}. The vanishing inverse temperature at $E=E_m$ corresponds to a homogeneous state with positive and negative vortices spread out evenly over the domain. The point-vortex states at different temperatures are summarized in figure \ref{fig:overview_vortices}. \NOTE{Such point-vortex states at any given inverse temperature $\beta$ may be generated using the noisy gradient method presented in appendix \ref{sec:appA}, which was previously introduced in \cite{krstulovic2009generation}. Specifically, once a statistically stationary state is reached, this numerical method generates random point-vortex states according to the canonical distribution associated with the inverse temperature $\beta$. For a given value of $\beta$, the mean energy in the statistically stationary state can be measured from the time series. Thus, like every microcanonical temperature corresponds to an energy $E$ according to (\ref{eq:def_beta}), in the noisy gradient method every value of $\beta$ corresponds to a mean energy $\langle E \rangle$ in steady state. The resulting mean energy as a function of temperature is shown in figure \ref{fig:E_vs_beta}.}
%%%%%%%%%%%%%%%%%%%%%%%%%%%%%%%%%%%%%%%%%%%%%
%%%%%%%%%%%%%%%%%%%%%%%%%%%%%%%%%%%%%%%%%%%%%

\section{The model}             % ALEX             %%%
\label{sec:model}                                  %%%
%%%%%%%%%%%%%%%%%%%%%%%%%%%%%%%%%%%%%%%%%%%%%%%%%%%%%%
Here we \NOTEE{construct} the simplified model of the interaction of 2-D and 3-D flow studied in this paper. 
\textcolor{black}{
The model is in the same spirit as shell models of turbulent cascade processes \cite{biferale2003shell}, which replace the Navier-Stokes dynamics with a simpler set of coupled nonlinear ordinary differential equations, which conserve a number of quantities including total energy and enstrophy in the 2-D case, aiming at providing insights into turbulent cascade processes. The present model, as we show below, may similarly provide insights into the dynamics of 3-D instabilities on turbulent  2-D flows.}

For the sake of simplicity and clarity, the theoretical formalism is presented in the infinite domain. In appendix \ref{sec:app_model_in_periodic_bc}, we provide the equations for the 2-D doubly periodic domain $[0,2\pi L]\times [0,2\pi L]$, where the statistical point-vortex temperature from section \ref{sec:bg_statmech} is well defined.

\NOTEE{Our main goal is to arrive at a model of minimum complexity describing the growth of 3-D perturbations on a 2-D large-scale condensate flow. Two key ingredients must be selected.}
\NOTEE{Firstly, a model of the two-dimensional base flow must be chosen. Here we opt for 2-D point-vortex flow, in view of its many successful modelling applications to two-dimensional turbulent flows, as presented in the introduction. Specifically, we consider an even number $N_v$ of point vortices with circulations $\Gamma_i=\Gamma$ for odd $i$ and $\Gamma_i=-\Gamma$ for even $i$, located at positions $\mathbf{x}_v^{(i)}=(x_v^{(i)},y_v^{(i)})$.}

\NOTEE{Secondly, the 3-D perturbations have to be modelled.  While there exist 3-D vortex filament models, commonly used in quantum turbulence, which describe mutual advection of curved vortex lines  \cite{bustamante2015derivation,hanninen2014vortex}, these are significantly more complex than their 2-D counterparts -- in particular, each segment of every vortex line is advected by all other vortex lines via the Biot-Savart law, and in addition proper handling of vortex reconnections is a complicating factor. Instead, here we seek a simpler description.
\textcolor{black}{Simulations of turbulent flows close to the onset of three-dimensionality reveal that 3-D perturbations are strongly localized (spatially intermittent) in the 2-D plane \cite{benavides2017critical,van2019condensates,seshasayanan2020onset}. Indeed,} close to the onset of three-dimensionality, high wavenumbers in the third dimension are suppressed by viscous damping. Hence, the 3-D instability, while being strongly localized in the 2-D plane, is also expected to have a simple spatial structure in the third dimension, and its intensity can be approximately characterized by a single scalar amplitude.}

\NOTEE{Combining these two insights, we model 3-D motions as $N_p$ localized, point-like entities in the plane whose detailed spatial structure in the third dimension is ignored, and whose intensity is characterized by an effective perturbation amplitude $A_k$, for $k=1,\dots,N_p$.  We name these entities ``{\it ergophages}" and denote their positions by $\mathbf{x}_p^{(k)}=(x_p^{(k)},y_p^{(k)})$. While the model describes 3-D flow, the mathematical structure of the model is effectively 2-D. We stress that this is not a contradiction, since the reduction is based on the physical properties of 3-D perturbations close to onset, and retains 3-D information.} 

%In addition to the point vortices, we introduce localised 3-D perturbations . 
%Motivated by the strong localisation of 3-D motions (spatial intermittency) observed close to the onset of three-dimensionality in simulations \cite{benavides2017critical,van2019condensates,seshasayanan2020onset} we also model 3-D motions as $N_p$ point-like pertubations
%\NOTE{ located at positions $\mathbf{x}_p^{(k)}=(x_p^{(k)},y_p^{(k)})$}.

Point vortices and 3-D perturbations induce velocity fields that advect each other following the equations
\begin{equation}
    \frac{d}{dt} \mathbf{x}_v^{(i)} =   \NOTE{{\mathbf{U}'}}_{v}^{(i)} + \mathbf{U}_{p}^{(i)} + \mathbf{u}_f^{(i)} \label{eq:xv_evolution_inf2}
\end{equation}
and
\begin{equation}
    \frac{d}{dt} \mathbf{x}_p^{\NOTE{(k)}} =   \mathbf{U}_{v}^{\NOTE{(k)}} + \mathbf{v}_f^{\NOTE{(k)}} \label{eq:xp_evolution_inf2}
\end{equation}
where \NOTE{$\mathbf{U}_{v}'^{(i)}$ is the velocity induced on vortex $i$ by all point vortices $i\neq j$}, $\mathbf{U}_{p}^{(i)}$ is the velocity induced on vortex $i$ by the 3-D {\it ergophages} and $\mathbf{U}_v^{(k)}$ is the velocity induced on ergophage $k$ by all $N_v$ point vortices. Finally, $\mathbf{u}_f^{(i)}$ and $\mathbf{v}_f^{(k)}$ are externally imposed velocity fields that could inject energy to the system. Also, note that ergophages do not advect each other, \NOTEE{a choice which is made for simplicity -- mutual advection of ergophages can easily be included in the model presented below (while this was not studied in detail, it did not seem to affect the qualitative model behavior). }

%\NOTEE{The fact that ergophages are 3-D structures implies that $\mathbf{U}_{p}^{(i)}$ has a non-zero divergence in the $(x,y)$ plane，as opposed to $\mathbf{U}_v^{(i)}$ and $\mathbf{U}_{v}^{(i)}$, whose divergence vanishes.} 

In the absence of ergophages and external velocities, \NOTEE{the model reduces to classical point-vortex flow. In this case,} point vortices move due to their mutual advection, following Hamiltonian dynamics so that the 
velocity field ${\mathbf{U}'}_{v}^{(i)}$
%\CMNT{$=U_v({\bf x}_{v,p}^i) $} 
can be written as
%\CMNT{
%\begin{equation}
%    {\mathbf{U}}_{v}({\bf x}) =\sum_j \begin{pmatrix}  \partial_y %\psi_j({\bf x}) \\ - \partial_x \psi_j({\bf x})\end{pmatrix},  
%    \label{eq:def_Uv2}
%\end{equation}
%where
%\beq
%\psi_i({\bf x}) = \Gamma_i \log(|{\bf x}_v^{(i)}-{\bf x}| ).
%\eeq
%}
\begin{equation}
    {\mathbf{U}'}_{v}^{(i)} = \NOTE{\Gamma_i^{-1}} \begin{pmatrix}  \partial_{y_v^{(i)}} H \\ - \partial_{x_v^{(i)}} H \end{pmatrix},  
    \label{eq:def_Uv}
\end{equation}
\NOTE{corresponding to the advection of the $i$-th vortex by all vortices $j\neq i$. The Hamiltonian $H$ in $\mathbb{R}^2$ is given by} 
\begin{equation}
    H (x_v^{(1)},\dots, x_v^{(N_v)} )= - \frac{1}{2} \sum_{i,j=1\atop i\neq j}^{N_v}  \Gamma_i \Gamma_j \log(|\mathbf{x}_v^{(i)} - \mathbf{x}_v^{(j)}|),
    \label{eq:def_H}
\end{equation}
which is a sum over pairs depending on the vortex-vortex distances alone. \NOTE{The velocity field $\mathbf{U}_v^{(k)}$ closely resembles ${\mathbf{U}'}_v^{(i)}$, but it includes the advection due to all $N_v$ vortices, formally omitting the condition $i\neq j$ in $H$ before differentiating in (\ref{eq:def_Uv}) and evaluating at $\mathbf{x}_v^{(i)}\to\mathbf{x}_p^{(k)}$.} The Hamiltonian also gives the \NOTE{kinetic} energy \NOTE{of the flow} (up to a factor of $(2\pi)^{-1}$ times the constant fluid density, and an additive infinite constant due to self-energy), which is conserved. The point-vortex energy increases when same-sign vortices approach each other and when opposite-sign vortices move apart, while it decreases when same-sign vortices move apart and when opposite-sign vortices approach each other.

In the presence of ergophages, energy of the 2-D field can be transferred to the 3-D field perturbations. Thus, in order to gain energy, an ergophage must reduce the energy of a given point-vortex configuration on which it is superimposed. Each ergophage induces a 3-D perturbation velocity field $\mathbf{u}^k_{p}(\mathbf{x})$ of amplitude $A_k^2$. \NOTEE{Importantly, despite the model being formally 2-D, the fact that ergophages represent 3-D structures implies that $\mathbf{u}_{p}^{(k)}(\mathbf{x})$ has a non-zero divergence in the $(x,y)$ plane. This is in contrast to the velocity field $\mathbf{U}_v^{(i)}(\mathbf{x})$ induced by 2-D point vortices, whose 2-D divergence vanishes.} The total velocity field induced by the ergophages is then given by
\begin{equation}
    \mathbf{U}_{p}(\mathbf{x})= \sum_{k=1}^{N_p} A_k^2 \mathbf{u}^{(k)}_{p}(\mathbf{x}),
    \label{eq:Up_tot}
\end{equation}
\NOTE{such that the velocity induced on vortex $i$ can be written as $\mathbf{U}_p^{(i)}=\mathbf{U}_p\left(\mathbf{x}_v^{(i)}\right)$.} This field modifies the point-vortex positions and thus their energy, allowing ergophages to grow under suitable conditions.

\begin{figure}
    \centering
    \includegraphics[width=6.5cm]{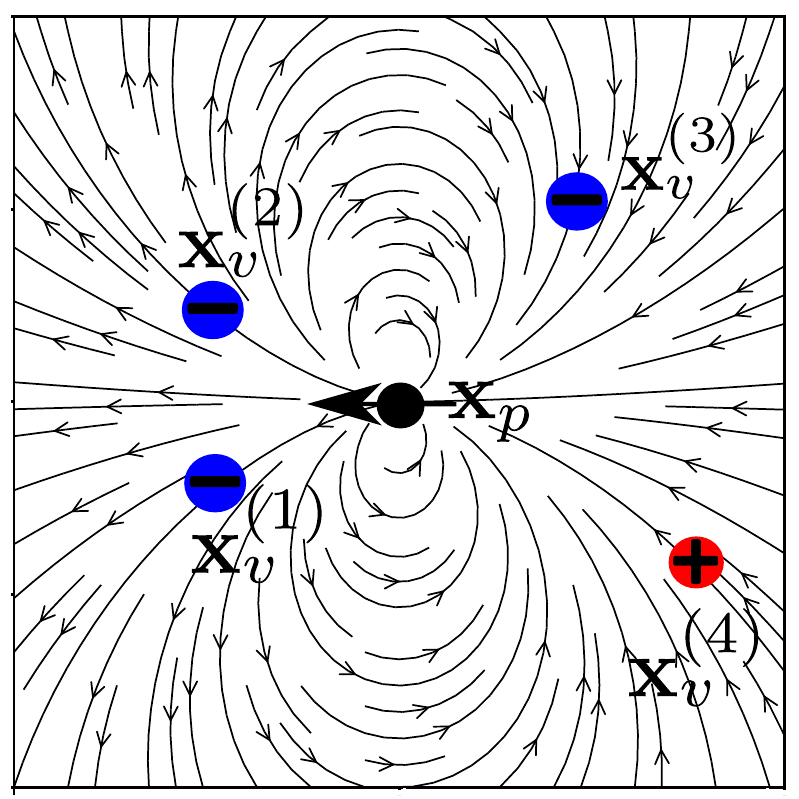}

    \caption{ Illustration of how a velocity field $\mathbf{u}_p$ (steam lines) due to a 3-D perturbation at $\mathbf{x}_p$, can reduce point-vortex energy. This is done by increasing the distance between the same-sign vortices at $\mathbf{x}_v^{(1)}, \mathbf{x}_v^{(2)}$ and/or decreasing the distance between opposite-sign vortices at $\mathbf{x}_v^{(3)},\mathbf{x}_v^{(4)}$. 
    %Two choices of $\mathbf{u_p}$ are shown: a monopole field (left panel). dipole field (right panel). 
    The bold black arrow passing through $\mathbf{x}_p$ represents the dipole moment. }
    \label{fig:monopole_dipole_ill}
\end{figure}

\NOTE{
Our choice for $\mathbf{u}^{(k)}_{p}(\mathbf{x})$ should be the simplest possible. \NOTEE{It is shown in the appendix \ref{sec:appC} that the choice of a monopole, which at first does suggest itself for its simplicity, cannot produce 3-D instability}}. Hence the simplest non-trivial choice for  $\mathbf{u}^{(k)}_{p}(\mathbf{x})$ is given by a dipole field,
%Two possible choices for $\mathbf{u}^{(k)}_p$ are shown in figures \ref{fig:monopole_dipole_ill} for a single ergophage, namely a monopole and a dipole field. In both cases the induced velocity field can be expressed in terms of a potential $\phi^{(k)}$, for the monopole as
%\begin{equation}
%    \mathbf{u}^{(k)}_{p} = \nabla \phi^{(k)} = \begin{pmatrix} \partial_x \phi^{(k)} \\ \partial_y \phi^{(k)} \end{pmatrix}
%\end{equation}
%and for the dipole
\begin{equation}
    \mathbf{u}^{(k)}_{p} = (\NOTE{\hat{d}_k} \cdot \nabla)  \begin{pmatrix} \partial_x \phi^{(k)} \\ \partial_y \phi^{(k)} \end{pmatrix}.
    \label{eq:dipole_field}
\end{equation}
\NOTE{where $\hat{d}_k=(\cos(\varphi_k),\sin(\varphi_k))$ is the dipole moment
with $\varphi_k$ the angle between the dipole moment and the $x$-axis.} The potential $\phi^{(k)}$ is given by 
\begin{equation}
    \phi^{(k)} (\mathbf{x} )= - \frac{1}{2} c   \log(|\mathbf{x}_p^{(k)} - \mathbf{x} |),
    \label{eq:def_phi}
\end{equation}
where $c$ is a coupling coefficient. 
An example of dipole interactions is shown in figure \ref{fig:monopole_dipole_ill}. In this case the perturbation velocity field makes same-sign vortices approach each other (e.g. $\mathbf{x}_v^{(1)}$ and $\mathbf{x}_v^{(2)}$ in figure \ref{fig:monopole_dipole_ill}) and opposite-sign vortices move apart (e.g. $\mathbf{x}_v^{(3)}$ and $\mathbf{x}_v^{(4)}$ in figure \ref{fig:monopole_dipole_ill}), thus reducing the point-vortex energy. Now, assume one were to interchange $\mathbf{x}_v^{(1)}\leftrightarrow \mathbf{x}_v^{(4)}$ and $\mathbf{x}_v^{(2)}\leftrightarrow \mathbf{x}_v^{(3)}$  in figure \ref{fig:monopole_dipole_ill}, keeping $\mathbf{x}_p$ the same. 
The dipole field would then cause an increase in point-vortex energy and thus would no longer lead to any 3-D instabilities. However, it suffices to rotate the dipole moment by $180^{\circ}$ to recuperate a 3-D instability. This example illustrates that 
%with a suitable evolution of the orientation of the dipole moment, 
the dipole field can lead to 3-D instability for a given vortex configuration (even if monopole field would not), provided that the orientation of the dipole moment is suitably chosen. 
For simplicity the dipole moment in this work will always be chosen such as to \NOTE{ensure maximum} \NOTEE{(positive)} energy extraction from the 2-D field.\\  
%%%%%%%%%%%%%%%%%%%%%%%%%%%%%
%--------------------------------------------------------------------
%--------------------------------------------------------------------
%%%%%%%%%%%%%%%%%%%%%%%%%%%
In our model we assign to the ergophages the 3-D energy 
\begin{equation} E_{3D} = \frac{1}{2}\sum_k A_k^2 . \end{equation} 
The energy exchanges between 2-D and 3-D flow must be conservative. 
Thus any decrease of the point-vortex energy should correspond to an increase 
of 3-D ergophage energy. We let the amplitudes $A_k$ evolve according to
\begin{equation}
    \frac{\mathrm{d} A_k}{\mathrm{d}t} =  (\gamma_k-\nu) A_k - \delta A_k^3  \label{eq:A_evolution_inf2}
\end{equation}
\NOTE{(no implicit summation)}, where $\gamma_k$ is an instantaneous growth rate
due to interactions with the point vortices, $\nu$ (proportional to viscosity) is a linear damping coefficient and $\delta$ is a nonlinear damping coefficient due to self-interactions. \textcolor{black}{Such nonlinear effects in three-dimensional velocity fields are associated with a Kolmogorov forward energy cascade, whose amplitude will generally depend on system parameters, such as domain geometry and system rotation rate. Hence the coefficient $\delta$ should also depend on these system parameters.}
In order for the coupling to conserve energy, the growth-rate is given by 
\begin{equation}
    \gamma_k= -  \sum_{i=1}^{N_v}  {\bf u}^{(k)}_p\left(\mathbf{x}_v^{(i)}\right) \cdot \nabla_{{\bf x}_v^{(i)}} H .
    \label{eq:defgamma_inf}
\end{equation}
As is shown in appendix \ref{sec:appB}, these model equations %\NOTEE{the model equations (\ref{eq:xv_evolution_inf2}), (\ref{eq:def_Uv}), (\ref{eq:Up_tot}), (\ref{eq:A_evolution_inf2}) and (\ref{eq:defgamma_inf})}
imply that
% (independently of the choice of $\hat{\mathbf{u}}_p$ monopole or dipole) 
 the total energy
\begin{equation}
    E_{tot} = H + \frac{1}{2}\sum_{k=1}^{N_p} A_k^2 = H + E_{3D} \label{eq:Etot}
\end{equation}
is conserved, provided $\mu=\delta=0$ (no dissipation) and $\mathbf{u}_f=\mathbf{0}$ (no energy injection). \NOTEE{Note that for $E_{tot}$ to be dimensionally consistent, $A_k$ must have dimensions of circulation. In addition to the energy, the 2-D Euler equation conserves the so-called Casimir invariants, which are of the form $\int \omega^n d^2x$, ($n=2$ gives the enstrophy), where $\omega$ denotes vorticity. In the point-vortex model, the vorticity depends only on the number and circulation of vortices, both of which are conserved in our model. }

In the presence of dissipation it is useful to have a \NOTEE{driving mechanism} as well, so that a non-trivial steady state is reached.
This is achieved by the choice
%, we \NOTE{switch on the "driving" term} on the right-hand side of the point vortex equation of motion (\ref{eq:xv_evolution_inf}), given by a constant $\epsilon_f$ times the noisy gradient dynamics described in appendix \ref{sec:appA},
    \begin{equation}
        \mathbf{u}_f^{(i)} =  \epsilon_f\left[ \nabla_{\mathbf{x}_i} H + |\beta_f|^{-1/2} \boldsymbol{\eta}_i(t)\right], \label{eq:uf}
    \end{equation}
where $\boldsymbol{\eta}_k(t)= (\eta_k^1(t),\eta_k^2(t))^T$ with independent white Gaussian noise components $\eta_k^i$ satisfying $\langle \eta_k^i (t) \rangle =0$ and $\langle \eta_k^i \eta_{k'}^{i'}\rangle = 2 \delta_{i,i'} \delta_{k,k'}\delta(t-t')$ for the ensemble average $\langle\cdot \rangle$. 
In the absence of ergophages, this noisy-gradient driving leads to a point-vortex flow with temperature $\beta_f^{-1}$ and is described in detail in appendix \ref{sec:appA}. \NOTEE{We emphasize that the driving (\ref{eq:uf}) can either increase or decrease the 2-D energy. If the 2-D energy at any given time is above the equilibrium value corresponding to the temperature $\beta_f^{-1}$ (shown in Fig. \ref{fig:E_vs_beta}), then the driving will act to decrease energy to the equilibrium value. Conversely, if the 2-D energy is below that equilibrium value, the driving will act to increase the 2-D energy. We also point out that, as a consequence of the inverse energy cascade, 2-D flows typically feature the formation of large-scale coherent structures at late times. Such a structure is observed in the point-vortex system at negative $\beta$. At intermediate stages of the inverse cascade process, for instance if the cascade is interrupted by large-scale friction, one finds an approximately homogeneous gas of vortices \cite{mcwilliams1984emergence}. In the point-vortex system, this is realized when $\beta\approx 0$. At $\beta>0$, the point-vortex model is characterized by vortex-antivortex bound states. To the best of our knowledge, however, these are never observed in laboratory experiments \cite{xia2011upscale,kellay2017hydrodynamics} nor numerical studies \cite{celani2010turbulence,benavides2017critical,van2019condensates} of turbulent quasi-2-D flows. We conclude that the regime $\beta\leq0$ is the physically relevant one.}
%i.e. $\langle \eta^{(1)}_i\rangle = \langle \eta^{(2)}_i\rangle  =0$ and $\langle \eta^{(j)}_i(t) \eta^{(j')}_{i'}(t') \rangle = 2 \delta(t-t') \delta_{i,i'} \delta_{j,j'}$, in terms of the ensemble average $\langle \cdot \rangle$.
%We choose $\beta_f<0$ with $|\beta_f|$ large, such that in the absence of 3-D perturbations, $\mathbf{u}_f$ drives the system towards a condensate state.

Finally, since the total energy is independent of the ergophage positions, we chose ${\bf v}_f$ to be a noise term, without altering the energy dynamics, 
\begin{equation}
    {\bf v}^{(k)}_f = \sigma \boldsymbol{\eta}_k(t) \label{eq:def_vf}
\end{equation}
where $\mathbf{\eta}_i=(\eta_i^{(1)},\eta_i^{(2)})$, with $\eta_i^{(j)}$ pairwise independent zero-mean white Gaussian noise terms.
The noise is added to eliminate a remaining dependence on initial conditions. Note that in our model, different ergophages do not directly affect each other, neither in terms of their amplitudes, nor their positions. They can only affect each other indirectly by altering the background 2-D flow non-negligibly and thus changing the growth rate $\gamma_k$ experienced by each ergophage. \NOTEE{This is mainly motivated by our goal of maximum simplicity. Firstly, the model 3-D energy is independent of ergophage positions, thus we may decide to neglect mutual advection of ergophages in a minimal description of how 3-D energy evolves. Secondly, while in a strongly 3-D flow, the 3-D components of the flow will feed back on one another, the growth or decay of 3-D perturbations at small to moderate 3-D amplitudes on a primarily 2-D flow should be mainly determined by direct interactions between 2-D and 3-D components, rather than interactions between 3-D and 3-D components. }

Equations (\ref{eq:xv_evolution_inf2},\ref{eq:xp_evolution_inf2},\ref{eq:A_evolution_inf2}) define \NOTEE{the time evolution of} our model, which we solve numerically in the following sections.
\section{Numerical implementation} % and methodology}    %%
\label{sec:method}                                      %%
%%%%%%%%%%%%%%%%%%%%%%%%%%%%%%%%%%%%%%%%%%%%%%%%%%%%%%%%%%
%Here we describe how the model equations were numerically implemented and how we proceeded in the analysis of the model.
%\subsection{Numerical implementation}
We developed a fully MPI-parallelized Fortran program, using a fourth-order Runge-Kutta time stepper, to simulate the model in the 2-D doubly periodic domain $[0,2\pi L]\times [0,2\pi L]$, based on the Weiss-McWilliams formalism introduced in \cite{weiss1991nonergodicity}. \NOTE{The parallelization is implemented by assigning a subset of vortex-vortex pairs and vortex-ergophage pairs to each processor, over which to sum when computing quantities involving such pairs such as $\mathbf{U}_v^{(i)}, \mathbf{U}_p^{(i)}, H$ and $\gamma_k$.} The specific model equations for the periodic domain are given in appendix \ref{sec:app_model_in_periodic_bc}. Since the periodic domain has a finite area, the statistical point-vortex temperature introduced in section \ref{sec:bg_statmech} is well defined here and no vortices can escape to infinity. A regularization was introduced at distances smaller than a positive cut-off $\epsilon\ll 2\pi L$ (we set $\epsilon/(2\pi L)=0.015$), similarly as in \cite{krasny1986desingularization}. \NOTEE{This regularization is required to avoid blow-ups, i.e. events where the time step required by the CFL condition \cite{courant1928partiellen} for well-resolvedness becomes extremely small. The way the cut-off is introduced approximately corresponds to smearing out the delta-peaked vorticity over a circular patch of constant vorticity, also known as a {\it Rankine vortex} \cite{acheson1990elementary}. In a realistic turbulent flow, there is a cut-off at small length scales related to viscosity. We note that vortex merging does not occur in the point-vortex model used here, with or without a cut-off (however, it may be added explicitly as in \cite{carnevale1991evolution, benzi1992simple,weiss1993temporal,trizac1998coalescence,weiss1999punctuated}).}  \NOTE{The time step $\Delta t$ for the Runge-Kutta scheme is dictated by the maximum growth rate $\gamma_k$, which is associated with close encounters where some distances are of the order of $\epsilon$. For highly condensed configurations, where $N_v/2$ vortices form a cluster for each sign of circulation, each cluster comprises approximately $N_v^2/8$ vortex pairs contributing to $\gamma_k$. At small distances $ \left|\mathbf{u}^{(k)}_p\right|=O(\epsilon^{-2})$ and $\left|\nabla_{\mathbf{x}_v^{(i)}}H\right|=O(\epsilon^{-1})$, such that the time step thus bounded above by 
\begin{equation}
   \Delta t \lesssim (\max(\gamma_k))^{-1} \propto \frac{8\epsilon^3}{ N_v^2}. \label{eq:delta_t}
\end{equation}
For dilute vortex configurations, the largest growth rates stem from encounters between a single ergophage and a single vortex, such that $\Delta t \lesssim \epsilon^3$. This strong dependence of the required time step on the cut-off $\epsilon$, and the number of vortices $N_v$ for dense configurations, is an important limiting factor in terms of computational cost. The operation of the highest numerical complexity at every time step is the evaluation of $\gamma_k$, since it requires summing $O(N_v^2)$ vortex-vortex pairs for every $k=1,\dots,N_p$.}

%\subsection{Method of analysis}
%%%%%%%%%%%%%%%%%%%%%%%%%%%%%%%%%%%%%%%%%%%%%%%%%%%%%%%%%%
\section{Simulation results}                            %%
\label{sec:simres}                                      %%
%%%%%%%%%%%%%%%%%%%%%%%%%%%%%%%%%%%%%%%%%%%%%%%%%%%%%%%%%%

To study the model introduced in section \ref{sec:model}, we first use the noisy gradient method described in appendix \ref{sec:appA} to generate point-vortex states with $N_v=32$ vortices at both positive and negative temperatures.  \NOTE{This relatively small number of vortices is chosen in order to be able to run simulations for long times in order to obtain satisfactory statistics.} The energy of the resulting equilibria as a function of their inverse temperature $\beta$ is as shown in figure \ref{fig:E_vs_beta}. We note that at this relatively low number of vortices, the transitions to a condensate and to pair condensation are not sharp. 
%The associated threshold temperatures $\beta_c$, $\beta_{pc}$ are thus not
Using these states generated by the noisy gradient method as initial conditions for the point vortices, we proceed in the three following steps:
\begin{enumerate}
    \item[(A)] The passive, linear regime: perturbation amplitudes $A_k/\Gamma\ll 1$ and $\delta\to 0$ for a given background point-vortex flow. In this limit, the evolution equation (\ref{eq:A_evolution_inf2}) of $A_k$ is linear  and the point-vortex energy $H$ is constant in time since $\mathbf{U}_p= O(A_k^2)$ is negligible with respect to the conservative Hamiltonian advection terms.
    To investigate this limit we set ${\bf U}_p=0$ in (\ref{eq:xp_evolution_inf2}) and $\delta=0$ in (\ref{eq:A_evolution_inf2}).
    Since there is no dissipation in the system we also set ${\bf u}_f=0$.
    %To ensure these two properties, we numerically set $c=0$, while $c'>0 0$ and $\delta=0$.
    %
    %
    \item[(B)] The passive, nonlinear regime: still $A_k/\Gamma\ll 1$, such that $H$ still remains unaffected by the 3-D instabilities, but we include saturation of the amplitude $A_k$ due finite $\delta$, i.e. nonlinear self-interaction (in both the linear and passive nonlinear regimes, individual 3-D perturbations evolve independently). In this limit 
    ${\bf U}_p={\bf u}_f=0$ in (\ref{eq:xp_evolution_inf2}) as well.
    \item[(C)] The fully nonlinear regime, where the amplitudes $A_k/\Gamma=O(1)$, thus the induced ergophage velocity ${\bf U}_p$ is finite and its effect on point vortices cannot be neglected. In this case $H$ is no longer \NOTEE{conserved}. To sustain the dynamics against dissipation, the ``driving" term ${\bf u}_f$ given in eq. (\ref{eq:uf}) is included.
    
\end{enumerate}

%%%%%%%%%%%%%%%%%%%%%%%%%%%%%%%%%%%%%%%%%%%%%
\begin{figure}
    \centering
    \includegraphics[width=8.6cm]{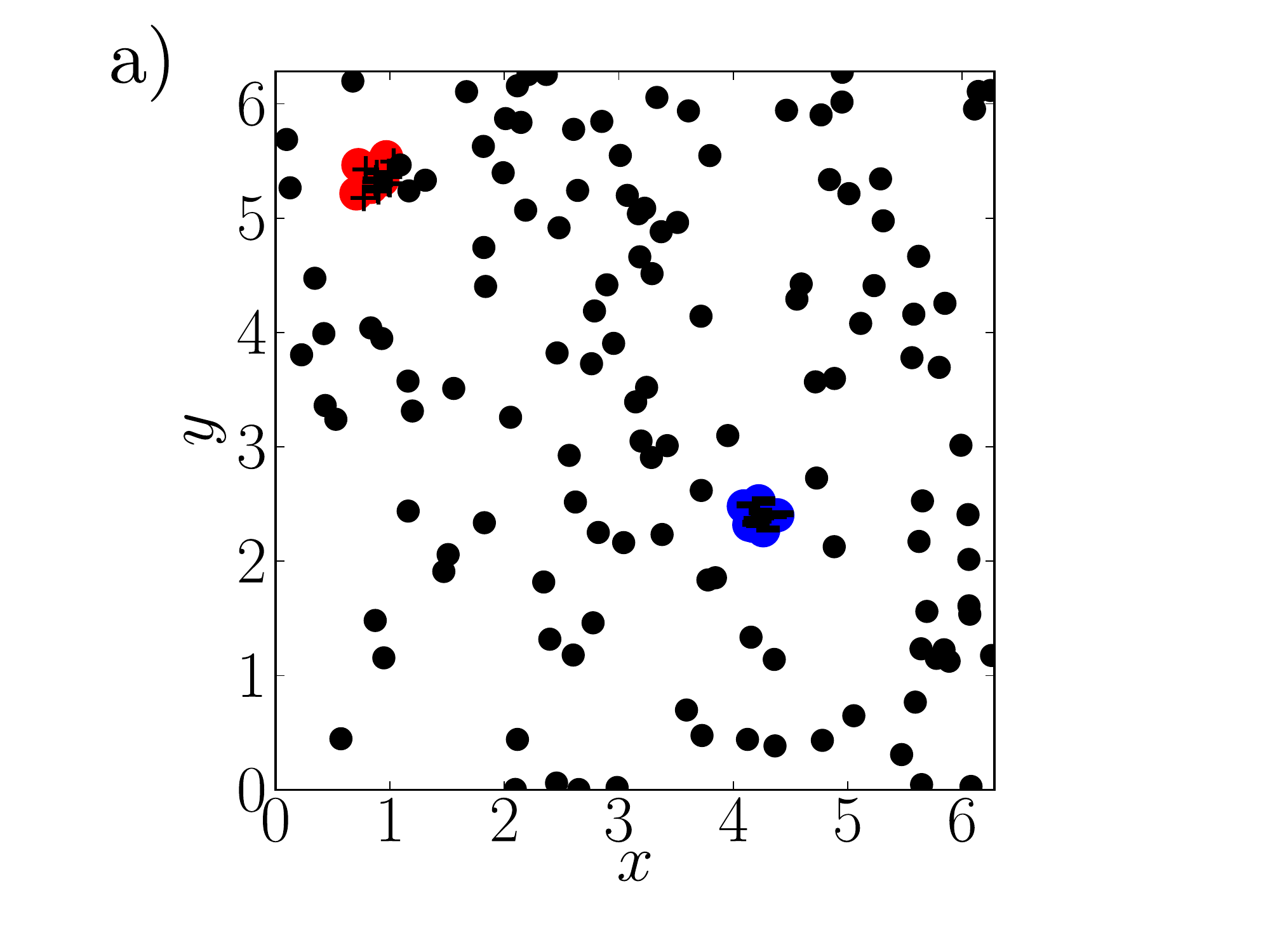}
    \includegraphics[width=8.6cm]{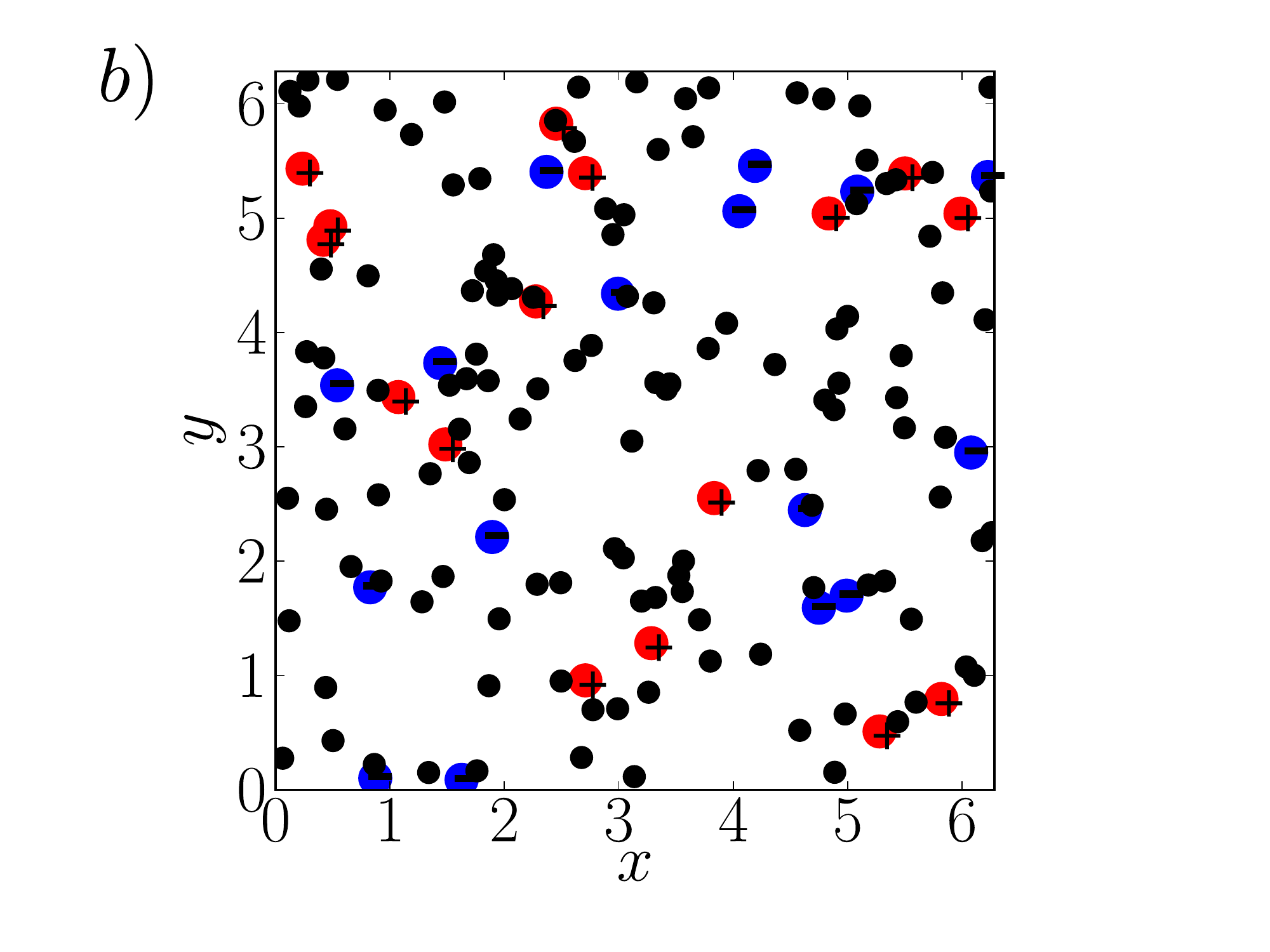} 
    \caption{Snapshots from two simulations with $N_p=128$ perturbations (black dots) evolving on a point-vortex flow consisting of  $N_v=32$ individual vortices, which is highly condensed at $\beta=-\frac{1}{8}$ (top) and dilute at $\beta=- \frac{1}{128}$ (bottom).}
    \label{fig:visualisaiton_T-8}
\end{figure}
   %%%%%%%%%%%%%%%%%%%%%%%%%%%%%%%%%%%%%%%%%%%%
   \begin{figure}
   \centering
   \includegraphics[width=8.6cm]{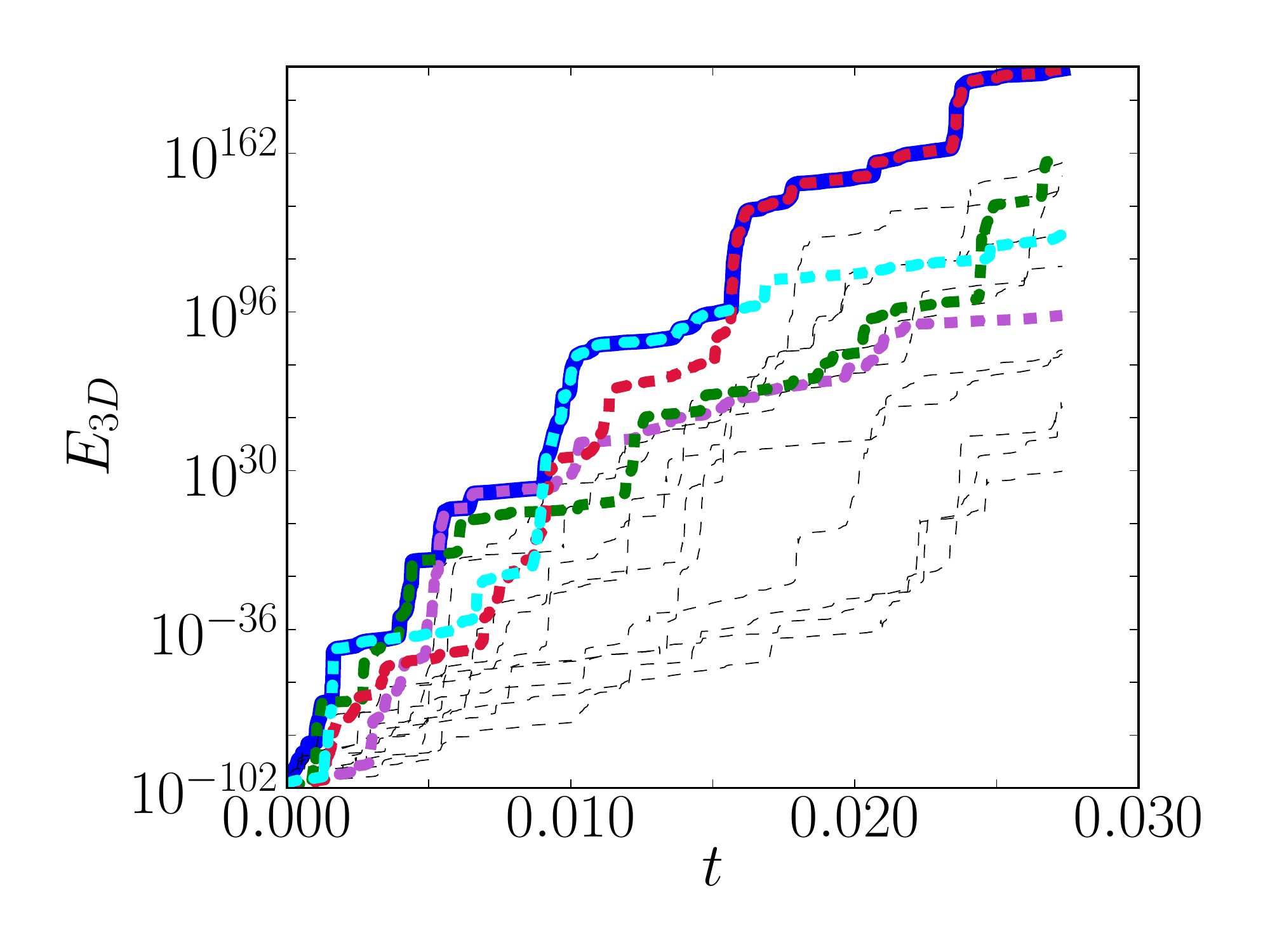}   
   \caption{Lin-log plot showing the time series of the energy of $N_p=128$ localized 3-D perturbations (total energy shown in solid blue, selected individual contributions $\frac{1}{2} A_k^2$ in dashed black lines) in the passive linear regime with $\nu=0$, growing on a highly condensed background 2-D flow at $\beta=-\frac{1}{8}$.}
   \label{fig:ts_-8}
   \end{figure}
   %%%%%%%%%%%%%%%%%%%%%%%%%%%%%%%%%%%%%%%%%%%%%

%Here we describe the results of our numerical simulations for the three different regimes under investigation.
%%%%%%%%%%%%%%%%%%%%%%%%%%%%%%%%%%%%%%%%%%%%%
\subsection{The passive linear regime}
\label{sec:simres_lin}
%%%%%%%%%%%%%%%%%%%%%%%%%%%%%%%%%%%%%%%%%%%%%
%  \subsubsection{Time evolution and growth rate statistics}
%  \label{sec:simres_lin_evolution_growth_rates}
%%%%%%%%%%%%%%%%%%%%%%%%%%%%%%%%%%%%%%%%%%%%%
   We initialize the simulation with $N_v=32$ vortices at an inverse temperature $\beta< 0$, with half of the vortices having circulation $\Gamma_i=\Gamma$, and the other half having circulation $\Gamma_i=-\Gamma$. In addition, we introduce $N_p=128$ randomly placed ergophages of some small initial amplitude (the same for every perturbation). It is worth \NOTE{reiterating} that in the linear phase of the evolution, since there is no feedback on the flow, each ergophage is evolving independently from all the others. Furthermore in the linear phase the effect of the damping parameter $\nu$ is to induce a mean exponential decay. The time evolution of $A_k(t,\nu)$ for any value of $\nu$ can thus be recovered from  the $\nu=0$ case as  $A_k(t,\nu)=A_k(t,0) e^{-\nu t}$. For this reason only the $\nu=0$ case is examined and the growth rate $\gamma_k'$ of a $\nu\ne0$ case is obtained as $\gamma_k'=\gamma_k-\nu$.
   
  The configuration under investigation is illustrated in figure \ref{fig:visualisaiton_T-8} for a highly condensed case ($\beta=-\frac{1}{8}$) and a  dilute case ($\beta=-\frac{1}{128}$). Then we let the system evolve in time and obtain a time series like the one shown in figure \ref{fig:ts_-8} for the highly condensed case, where the 3-D energy (solid blue line) alternates between plateau-like phases of slow growth and phases of abrupt exponential growth. The time series bears resemblance to that obtained from the complete linear stability analysis of 3-D instabilities on a turbulent 2-D flow performed by Seshasayanan and Gallet (see fig. 1 in \cite{seshasayanan2020onset}). In the same figure \ref{fig:ts_-8}, we also show the energy of individual ergophages, $\frac{1}{2}A_k^2$, by dashed lines. Their sum is equal to the blue solid line. 
  
  Two points need to be made. Firstly, one observes in the time evolution of individual ergophages that there are \NOTEE{alternating} phases of slow growth/stagnation and of rapid exponential growth. Secondly, at a given time $t$, $E_{3D}(t)$ is dominated by the ergophage with the largest amplitude $A_k(t)$. Abrupt growth events in $E_{3D}$ also occur when another ergophage $A_{k'}$ grows exponentially and ``overtakes" $A_k$, thereby leading to abrupt growth of the sum. 
  %   \CMNT{Adrian: I'm making a figure like fig. 6 for the monopoles.}   %------------------------------------------------------------------------------------
  %   \begin{figure} 
  %       \centering 
  %       \includegraphics[width=8.6cm]{time_series_monopole_decay.pdf} 
  %       \caption{Time series of mean and log-mean amplitude of monopole ergophages.}
  %       \label{fig:my_label}
  %   \end{figure}
  %------------------------------------------------------------------------------------
    \NOTEE{Each of the $N_p$ localized perturbations experiences a different, time-varying growth rate $\gamma_k(t)$. To understand this linear growth, we need to quantify the statistical properties of these random growth rates.}  
    
\begin{figure}
    \centering
    \includegraphics[width=8.6cm]{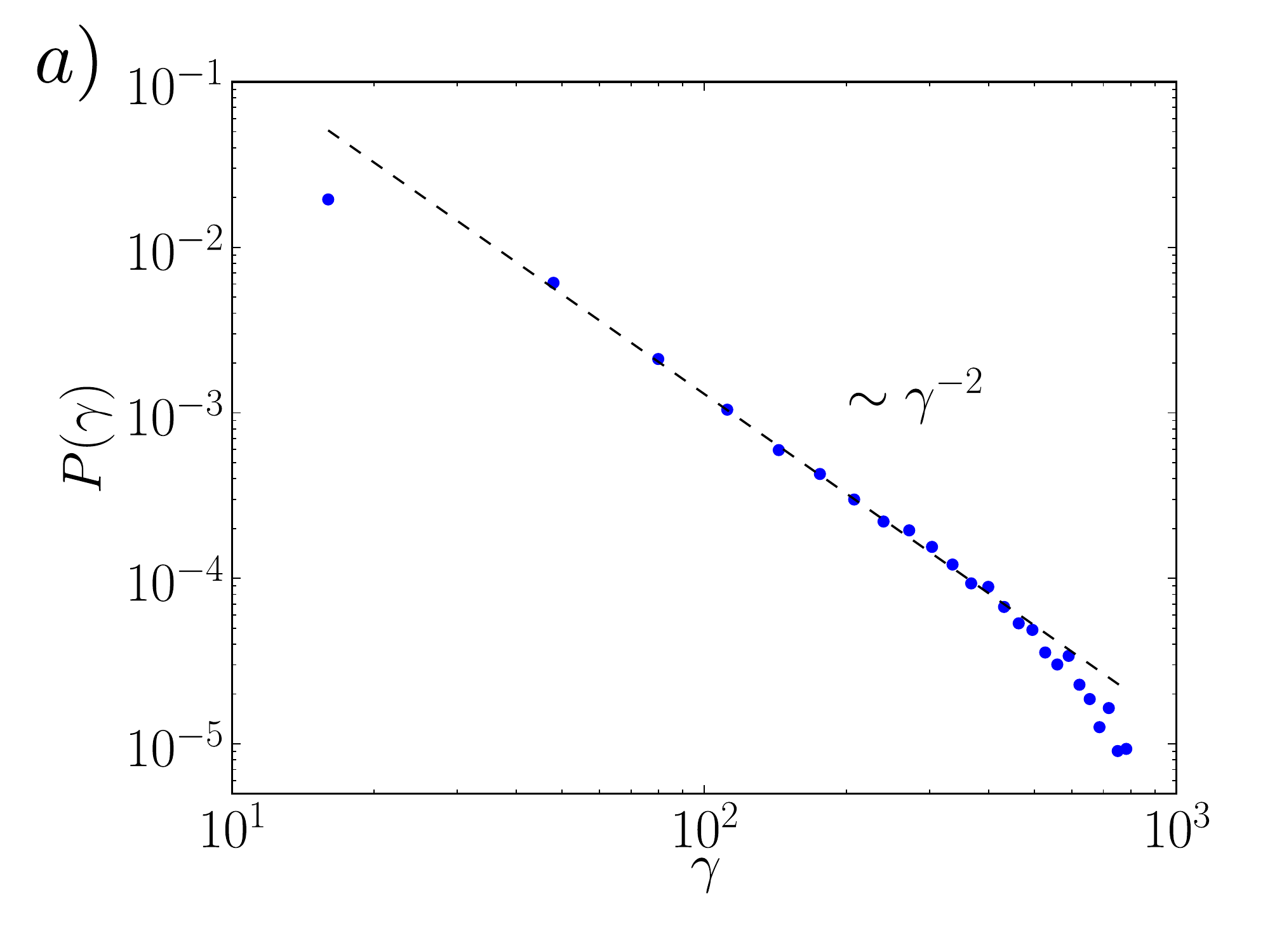}
        \includegraphics[width=8.6cm]{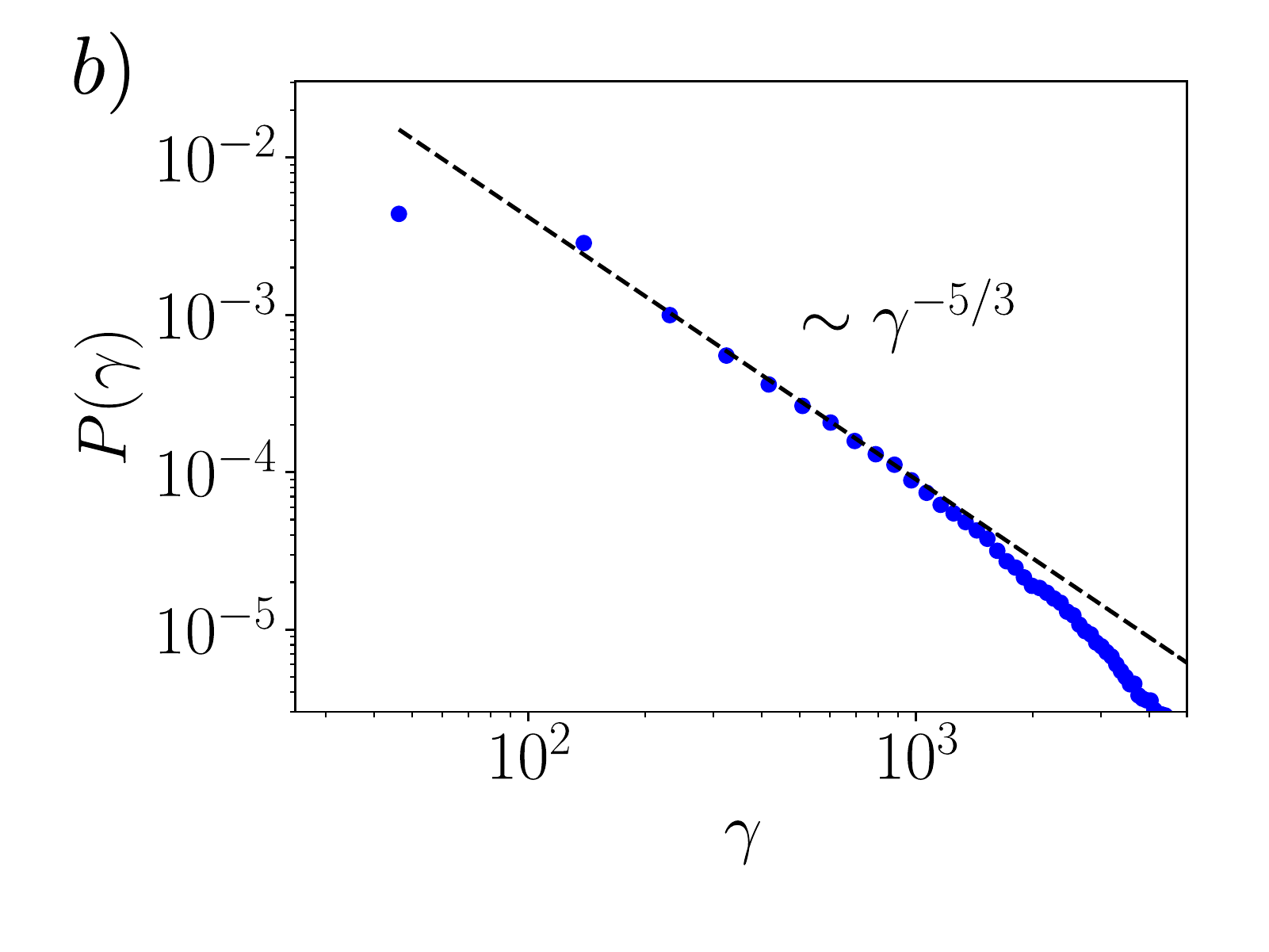}
    \caption{Two histograms of the growth rate $\gamma_k$, sampled over all time steps and all $128$ ergophages from the run corresponding to the two linear simulations with a dilute vortex state at $\beta=-1/128$ (top) and a condensed vortex state at $\beta=-1/8$ (bottom) visualized in figure \ref{fig:visualisaiton_T-8}. Power-law ranges with exponents $-2$ and $-5/3$ can be discerned, as predicted for dilute and dense vortex base states, respectively.}
    \label{fig:PDFs_gamma} 
   \end{figure}

    In figure \ref{fig:PDFs_gamma}, we plot histograms of $\gamma_k$ sampled over all $k=1,\dots,N_p$ and all time steps. In both cases, one observes a power-law range in the PDF. For the dilute case ($\beta=-\frac{1}{128}$) the power-law exponent is close to $-2$ while for the dense state ($\beta=-\frac{1}{8}$) it is closer to $-5/3$. These two exponents can be understood if one identifies the dominant interactions.
    In the dilute case $|\beta|\ll 1$, where point vortices are far apart, an ergophage maximizes its energy extraction when being close to a single point vortex. It does so by displacing the vortex towards the nearest opposite-sign vortex and/or further apart from the nearest same-sign vortex. 
    In the dense (condensate) case $-\beta\gg 1$, point vortices form high-density, same-signed clusters. In order for an ergophage to maximize energy extraction, it needs to be located close to these clusters. 
    % 
    %Similarly, in the case $\beta \gg 1$ \NOTE{(not demonstrated here)} where point-vortex-dipoles are formed, ergophages can only extract energy efficiently if they come close to such dipoles.
    %
    The PDF of the growth rate $\gamma_k$ can then be calculated by assuming that all positions in space are equally probable and that at each time it is the interaction with the closest pair of point vortices that dominates. A detailed calculation, given in appendix \ref{sec:appD}, yields %the power law 
    \begin{equation}
    \hspace{1cm}  P(\gamma) \propto \gamma^{-2}\hspace{1.2cm} \text{at large } \gamma.
    \end{equation}
    for the dilute limit $|\beta|\ll 1$, while for the dense %and point-vortex-dipole 
    limit $-\beta\gg 1$ one obtains \begin{equation}
    \hspace{1cm} P(\gamma)\propto \gamma^{-5/3} \hspace{1cm} \text{at large } \gamma.
    \end{equation}
    \NOTE{The predicted power laws agree with the PDFs obtained numerically.} Note, however, that in our numerical set-up these results are valid up to a large-$\gamma$ cut-off resulting from the regularization at distances less than $\epsilon$. This is important because without this regularization, the variance and the mean would be infinite for the power-law PDFs of $\gamma_k$ found here. This implies that some of the results observed here %would
    have an explicit dependence on $\epsilon$.
    %\NOTE{For the solutions of the Navier stokes equations this indicates that the stability of 2D flows would depend on the large scale dissipation mechanisms that limit the energy growth of the forming condensates.}
    %
    %in the dense case, which corresponds precisely to the theoretical predictions for $P(\gamma)$ presented in section \ref{sec:gamma_stats_and_Levy_flights}.
    %%%%%%%%%%%%%%%%%%%%%%%%%%%%%%%%%%%%%%%%%%%%%
   %%%%%%%%%%%%%%%%%%%%%%%%%%%%%%%%%%%%%%%%%%%%%
   %%%%%%%%%%%%%%%%%%%%%%%%%%%%%%%%%%%%%%%%%%%%%
   \begin{figure}
    \centering
    \includegraphics[width=8.6cm]{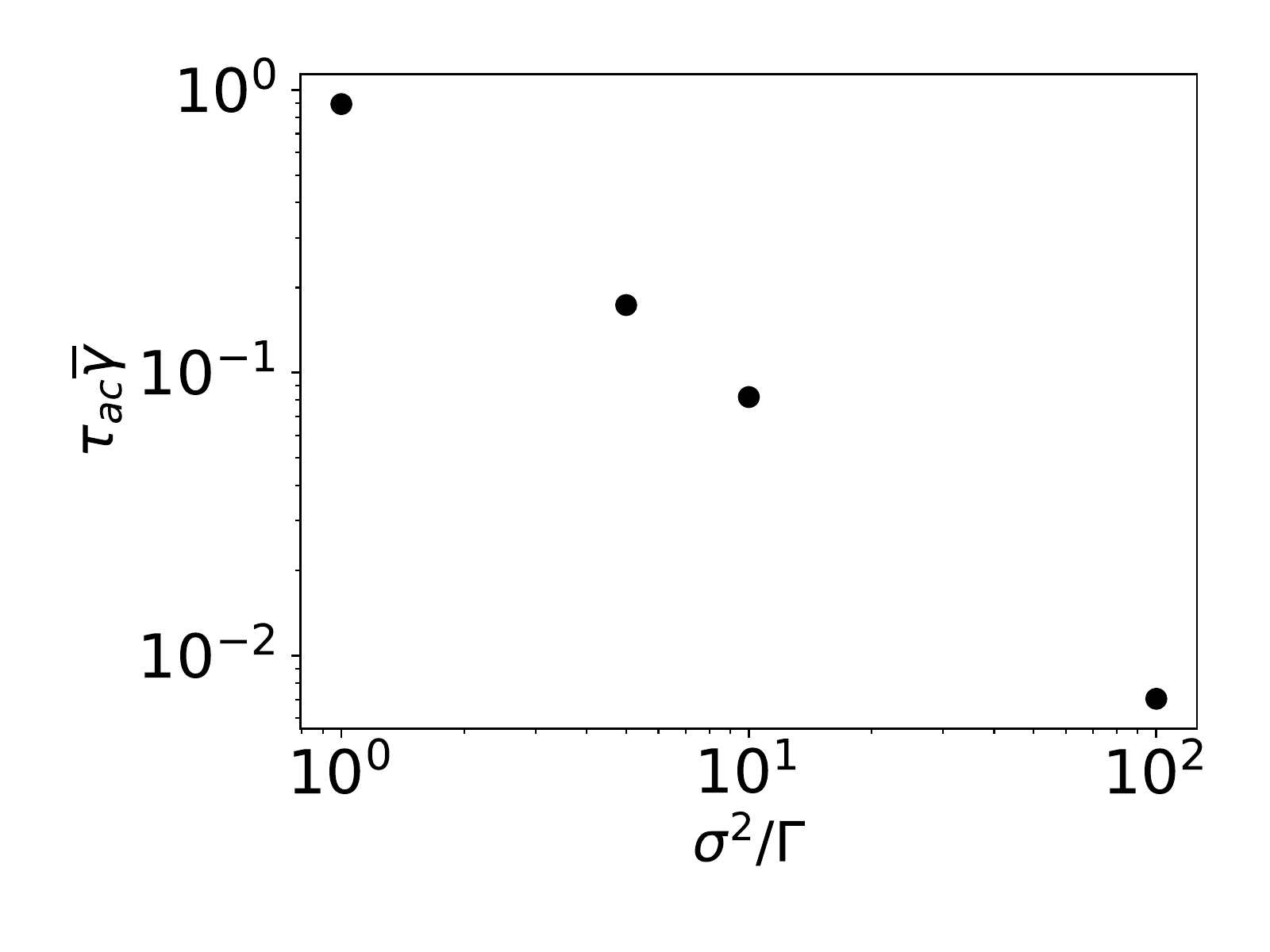}
    \caption{Log-log plot of the auto-correlation time $\tau_{ac}$ of the growth rate $\gamma$ (see text for the definition of $\tau_{ac}$) in a passive, linear simulation at $\beta=-1/16$, non-dimensionalized by the mean growth rate, versus the squared amplitude of the noise acting on 3-D perturbations, non-dimensionalized by the r.m.s. vortex circulation.}
    \label{fig:auto_correl_time}
    \end{figure}
    %--------------------------------------------------------------

   %%%%%%%%%%%%%%%%%%%%%%%%%%%%%%%%%%%%%%%%%%%%
   Besides the growth-rate distribution, to characterize the statistical properties of $\gamma_k$ 
   %We note furthermore that the fluctuating growth rate $\gamma$ has an 
   we also need to quantify its auto-correlation time $\tau_{ac}$.
   %which depends on the amplitude $\sigma$ of the Gaussian white noise acting on $\mathbf{x}_p$. 
   \NOTE{We define $\tau_{ac}$ in terms of the normalized auto-correlation function $\Gamma(\tau)=\langle \gamma(t) \gamma(t+\tau)\rangle/\langle \gamma(t)^2\rangle$, as the smallest $\tau$ for which $\Gamma(\tau)\leq 0.5$, where $\Gamma(0)= 1$ by definition and $\langle f(\gamma) \rangle = \frac{1}{N_pT} \int_0^T dt \sum_{k=1}^{N_p}  f(\gamma_k(t)) $ is an average over time $t$ ($T$ is the time at the end of the simulation) and realizations (ergophages).} We stress that the small-distance cut-off introduced in the velocity field, leading to a large-$\gamma$ cut-off in $P(\gamma)$ is essential for obtaining a finite mean growth rate $\langle \gamma \rangle$ and finite variance, since a PDF featuring power-law tails with exponents $-2$, $-5/3$ does not have a finite mean or variance otherwise. %, and hence $\Gamma(t)$ is not well-defined.
   Figure \ref{fig:auto_correl_time} shows that the auto-correlation time decreases monotonically with $\sigma$ (defined in (\ref{eq:def_vf})), as $\tau_{ac} \sim \sigma^{-2}$. By increasing $\sigma$ sufficiently, one obtains an arbitarily small auto-correlation time. When $\tau_c \langle \gamma \rangle \ll 1$, the random process $\gamma_k(t)$ can be approximated as uncorrelated in time. % When time is discretised, the growth rate  determines the steps of $\log(A_k)$.
   
   Summarising the above findings, the %length of these steps is
   \NOTEE{increments of $A_k$ are} randomly distributed according to a PDF with power law tails whose exponents are between $-2$ and $-5/3$ and approximately white in time since it is uncorrelated in time beyond a small correlation time (for sufficiently large $\sigma$). \NOTE{These properties imply that the evolution of $A_k$ due to $\gamma_k$ is well approximated by a L\'evy flight process.} 
    %A short introduction to such L\'evy flights is given in the following.
   
%%%%%%%%%%%%%%%%%%%%%%%%%%%%%%%%%%%%%%%%%%%%%%%%%%%%%%%%%%%%%%%%%%%%%%%%%%%%%%%%%%%%%%%%%%
% \subsubsection{L\'evy flights}
%%%%%%%%%%%%%%%%%%%%%%%%%%%%%%%%%%%%%%%%%%%%%%%%%%%%%%%%%%%%%%%%%%%%%%%%%%%%%%%%%%%%%%%%%%%
  A L\'evy flight is a random process with independent stationary increments $\eta$, where the increments follow a heavy-tailed PDF. \NOTE{By the generalized central limit theorem \cite{dubkov2008levy}, the sum of many such heavy-tailed increments follows a stable PDF} $P_{\alpha,\tilde{\beta}}(\eta)$ depending on two parameters
  $\alpha \in (0,2]$ and $\tilde{\beta}\in[-1,1]$.  
  L\'evy flights were first introduced in \cite{mandelbrot1983fractal} and have since found numerous applications in physics and beyond \cite{shlesinger1995levy,chechkin2008introduction}.  %\NOTEE{Bursts with amplitudes following power-law PDFs are also observed in 3-D secondary instabilities for zero-Prandtl-number convection \cite{kumar1996critical,kumar2006critical}.} %, including anomalous diffusion, in particular in fluid turbulence, \cite{shlesinger1987levy,solomon1993observation,metzler2000random,dubkov2008levy}, the statistics of 2-D fluid turbulence \cite{dubrulle1998truncated}, plasma turbulence \cite{del2005nondiffusive}, finance \cite{schinckus2013physicists}, climatology \cite{ditlevsen1999anomalous,ditlevsen1999observation}, human mobility \cite{rhee2011levy}, COVID-19 spreading \cite{gross2020spatio}, animal foraging patterns \cite{viswanathan1996levy} and more \cite{metzler2004restaurant, applebaum2004levy}.
  The influence of $\alpha,\tilde{\beta}$ on the PDF $P_{\alpha,\tilde{\beta}}(\eta)$ is as follows. For $\alpha=2$, one obtains the Gaussian distribution. For $\alpha<2$, a stable distribution features power-law tails $P(\eta)\propto \lbrace 1+\tilde{\beta}\mathrm{sign}(\eta)\rbrace  \eta^{-\alpha-1}$ at $|\eta|\to \infty$. The parameter $\tilde{\beta}$ measures the asymmetry of PDF. For $\tilde{\beta}=1$ and $\alpha<1$, one obtains a one-sided PDF \NOTE{with support on $\mathbb{R}_+$ only. \NOTEE{Stable PDFs are known to occur for velocity and velocity difference statistics in 2-D vortex flows in particular \cite{min1996levy}. The fact that the PDF of $\gamma_k$ shows power-law tails in our model can be understood as a consequence of this property of 2-D vortex flows.}}

If $\gamma_k$ is interpreted as noise, then equation (\ref{eq:A_evolution_inf2}) is a stochastic differential equation with multiplicative L\'evy noise \NOTE{whose parameters} depend on the 2-D flow temperature. 
The dense and dilute cases described above, for which the $\gamma_k$ PDF has power law ranges with exponents $-5/3$ and $-2$, respectively, correspond to noise parameters $\alpha=2/3$ and $\alpha=1$, respectively, and $\tilde{\beta}=1$ since the linear growth rate $\gamma_k$ is positive definite in the model by construction. 

The \NOTEE{theory of systems with multiplicative Gaussian white noise}
%in particular the It\^{o} and Stratonovich interpretations, are a standard part of courses on stochastic dynamics \cite{schenzle1979multiplicative,gardiner1985handbook,van1992stochastic}, and this 
has found a plethora of applications, in particular to noise-induced transitions \cite{horsthemke1984noise} and the phenomenon of on-off intermittency \cite{platt1993off,aumaitre2007noise,benavides2020multiplicative}. 
%If perhaps not as commonly taught in physics curricula, multiplicative L\'evy noise is the subject of a large number of studies in the physical and mathematical literature \cite{srokowski2009multiplicative,la2010dynamics,srokowski2010nonlinear,srokowski2012multiplicative,bessaih2015strong,zhai2015large}. 
While the role of long-time correlated noise in on-off intermittency has been considered before \cite{ding1995distribution, alexakis2009planar, alexakis2012critical, petrelis2012anomalous}, the case of on-off intermittency with heavy-tailed noise has not previously been studied explicitly, to our knowledge. Our companion paper \cite{vankan2020extreme} is devoted to this topic. Here we 
summarize only the relevant results. 
It is shown in \cite{vankan2020extreme} that \NOTE{in the case $\alpha<1$ and $\tilde{\beta}=1$, which applies here,} the system (\ref{eq:A_evolution_inf2}), with $\gamma_k$ interpreted as white L\'evy noise, is unstable for all values of $\nu$: since the mean value of $\langle \gamma_k\rangle \to +\infty$, viscosity $\nu$, no matter how large, cannot stop the growth of $A_k$. \NOTEE{If, however, the possible values $\gamma_k$ are restricted (``truncated") to be below some maximum}, so that a finite value of $\langle \gamma_k\rangle$ exists, then there is a critical value of viscosity $\nu_c$ above which all trajectories converge to zero $A_k\to0$. However, this critical value depends on the truncation value of $\gamma_k$, which implies that the threshold $\nu_c$ will depend on the regularization cut-off $\epsilon$. At long time scales the system displays on-off intermittency.  
%%%%%%%%%%%%%%%%%%%%%%%%%%%%%%%%%%%%%%%%%%%%%%%%%%%%%%%%%%%%%%%%%%%%%%%%%%%%%%%%%%%%%%%%%%%%%%%%%%%%%%%%%%%%%%%%%%%%%%

%%%%%%%%%%%%%%%%%%%%%%%%%%%%%%%%%%%%%%%%%%%%%%%%%%%%%%%%%%
\subsection{The passive nonlinear regime}
%%%%%%%%%%%%%%%%%%%%%%%%%%%%%%%%%%%%%%%%%%%%%%%%%%%%%%%%%%
%In contrast with free L\'evy flights, L\'evy flights with  noise in a steep potential $V(x)=\frac{a}{2}x^2+\frac{b}{4}x^4$, $a,b>0$, are known to exhibit finite variance even in the absence of a truncation \cite{chechkin2003bifurcation}. Here, in the passive nonlinear regime, where $\nu, \delta>0$ in eqn. (\ref{eq:A_evolution_inf}), we are faced with precisely this potential with multiplicative power-law noise.

\begin{figure}
    \centering
   \includegraphics[width=8.6cm]{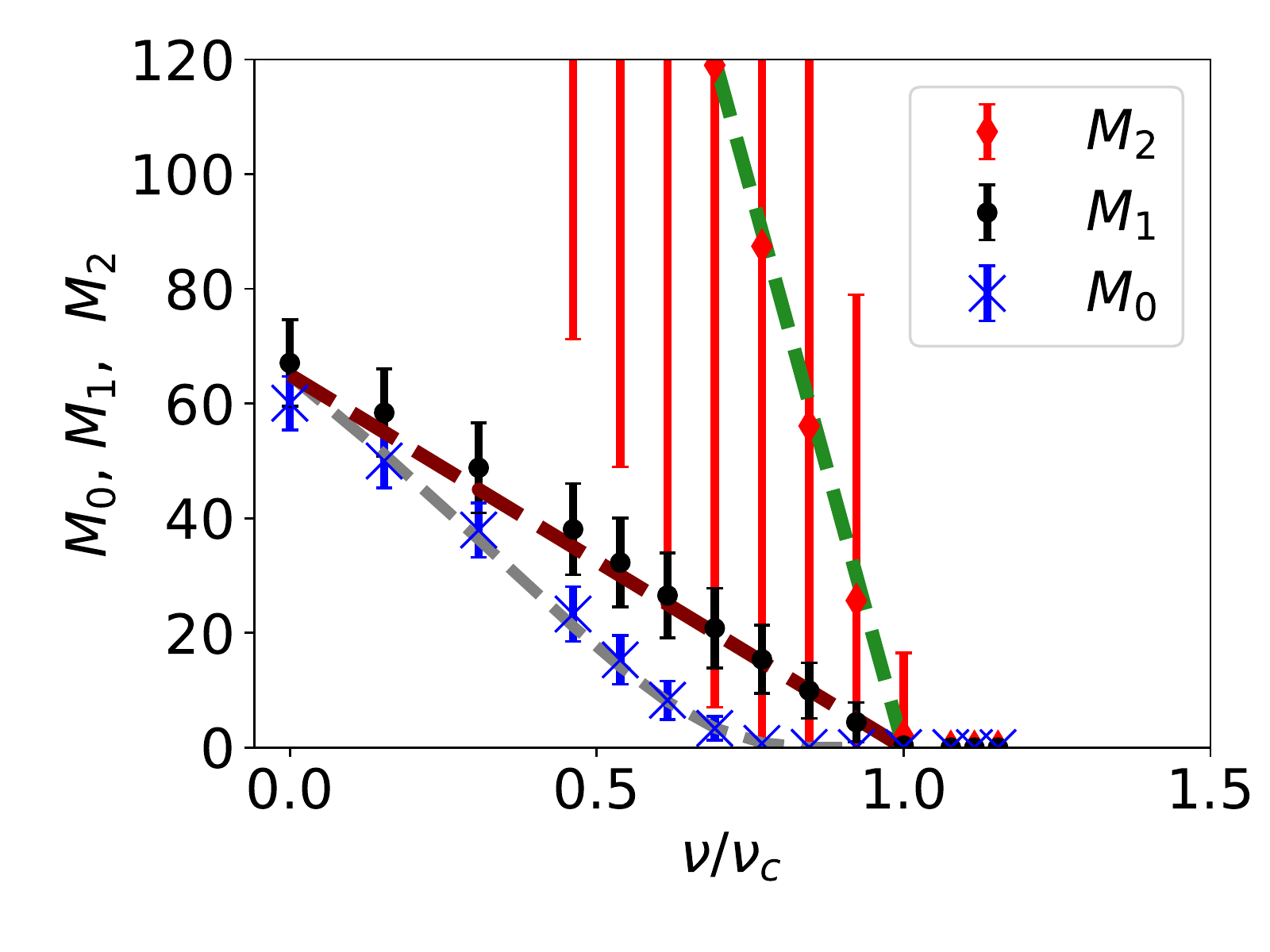}
    \caption{Bifurcation diagram for $N_p=32$ passive nonlinear (i.e. independent) dipole ergophages on the background flow at $\beta=-1/8$, $\delta=1$. The values of $\langle X \rangle_n$ is averaged over the statistically steady state. Error bars are given by the sample standard deviation of the time series in steady state. The dashed lines show the scalings from the Gaussian noise case.}
    \label{fig:bif_diag_T-8}
\end{figure}

We solve the model equations for $N_p=32$ passive nonlinear dipole ergophages evolving on a highly condensed background flow of $N_v=32$ point vortices at temperature $\beta=-1/8$, \NOTE{fixing the nonlinear damping coefficient at} $\delta=1$. For a given $\nu$, we initialize the ergophages at random positions and with small amplitudes. We let the system evolve for long times, such that the perturbation amplitude either decays or reaches a statistically steady state. We then measure the steady-state time average of the moments $M_n=\langle A^n \rangle$, in terms of $\langle f(A) \rangle =  \lim_{T\to \infty} \frac{1}{T N_p} \int_0^T \sum_{k=1}^{N_p} f(A_k)dt $. We also define the ``zeroth" moment as  $M_0=\exp(\langle \log(A)\rangle)$, %from $A^n \approx 1+n \log(A)+ \dots $, where the exponential is taken to obtain a positive definite quantity, as all other moments are positive. 
By the inequality of arithmetic and geometric means the moments are ordered $M_0\leq M_1\leq M_2^{1/2} \le M_3^{1/3} \leq  \dots$. The resulting bifurcation diagram of $M_0,M_1,M_2$ as a function of $\nu$ is shown in figure \ref{fig:bif_diag_T-8}.

On-off intermittency predicts that all non-zero moments scale 
linearly with $\nu_c-\nu$, $M_n\propto (\nu_c-\nu)$, while the zeroth moment scales as
$M_0\propto \exp(-cst./(\nu_c-\nu))$.
%Ordinary on-off intermittency is described, for instance, from the equation
%\begin{equation}
%    \dot{Y} = (\tilde{\eta}(t)+d)Y -b Y^3,
%\end{equation}
%which has the same form as eqn. (\ref{eq:Levy_flight_eqn_multiplicative}), but with $\tilde{\eta}(t)$ being Gaussian white noise with zero mean. There, for $0<d\ll 1$, i.e. close to onset of the linear instability, one finds $M_1,M_2\propto d$ , while $M_0\propto \exp(-cst./d)$.
%
Comparing this with \NOTEE{the bifurcation diagram shown in} figure \ref{fig:bif_diag_T-8}, where the scalings from the Gaussian case are shown by dashed lines, one sees that the time-averaged moments and the Gaussian scalings agree well within the errorbars. This is a consequence of the truncation in the model, which subjects the statistics to a convergence to the Gaussian case, albeit ``ultraslow" \cite{mantegna1994stochastic}, by the central limit theorem after the sample averaging and/or long-time averaging procedures.

%%%%%%%%%%%%%%%%%%%%%%%%%%%%%%%%%%%%%%%%%%%%%%%%%%%%%%%%%%
%%%%%%%%%%%%%%%%%%%%%%%%%%%%%%%%%%%%%%%%%%%%%%%%%%%%%%%%%%

%%%%%%%%%%%%%%%%%%%%%%%%%%%%%%%%%%%%%%%%%%%%%%%%%%%%%%%%%%
%%%%%%%%%%%%%%%%%%%%%%%%%%%%%%%%%%%%%%%%%%%%%%%%%%%%%%%%%%
%\begin{figure}
%    \centering
%   \includegraphics[width=7cm]{bif_th.pdf}
%    \caption{Bifurcation diagram obtained by averaging over $N_p=32$ independent realisations of the solution of eqn. (\ref{eq:Levy_flight_eqn_multiplicative}), where the parameters of the multiplicative L\'evy noise are $\alpha=2/3$, $\tilde{\beta}=1$, and the potential is $V(Y)= \nu Y^2/2 + \delta Y^2/4$, where we fix $\delta=0.001$.  Error bars are given by the sample standard deviation of the time series in steady state.}
%    \label{fig:bif_Levy_theory}
%\end{figure}
%%%%%%%%%%%%%%%%%%%%%%%%%%%%%%%%%%%%%%%%%%%%%%%%%%%%%%%%%%

%For comparison, we consider equation (\ref{eq:Levy_flight_eqn_multiplicative}) featuring multiplicative L\'evy noise with $\alpha=2/3$ and $\beta=1$, corresponding to the highly condensed vortex state at $\beta=-1/8$. We use a fourth order Runge-Kutta scheme with time step $\Delta t$ for the deterministic terms, $d\eta = \Delta t^{1/\alpha}$ for the noise, and truncating the noise at a large maximum value $\eta_{max}$ to obtain the diagram shown in figure \ref{fig:bif_Levy_theory}. The Gaussian scalings are again shown by the dashed lines and one observes once more that the steady-state, long-time-averaged moments exhibit the Gaussian scalings close to onset.

%%%%%%%%%%%%%%%%%%%%%%%%%%%%%%%%%%%%%%%%%%%%%%%%%%%%%%%%%%
\begin{figure}
    \centering
   \includegraphics[width=8.6cm]{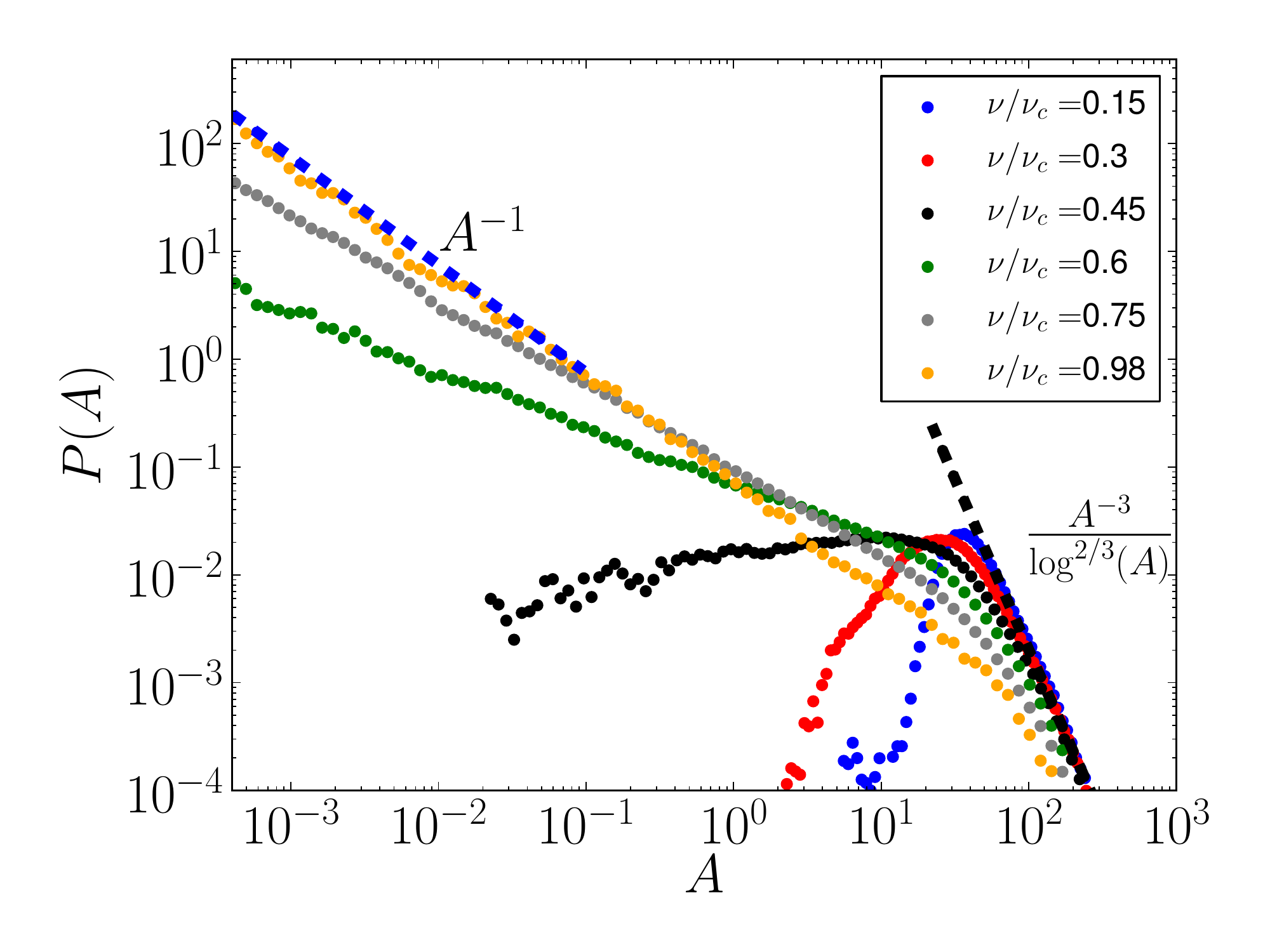}
    \caption{Steady state PDF of ergophage amplitudes from the numerical solution of the model in the passive nonlinear regime for $\delta=1$, $\nu/\nu_c=0.15$  on the background flow at $\beta=-1/8$.}
    \label{fig:PDF_steady_state}
\end{figure}
%%%%%%%%%%%%%%%%%%%%%%%%%%%%%%%%%%%%%%%%%%%%%%%%%%%%%%%%%%

%%%%%%%%%%%%%%%%%%%%%%%%%%%%%%%%%%%%%%%%%%%%%%%%%%%%%%%%%%
\begin{figure}
    \centering
   \includegraphics[width=8.6cm]{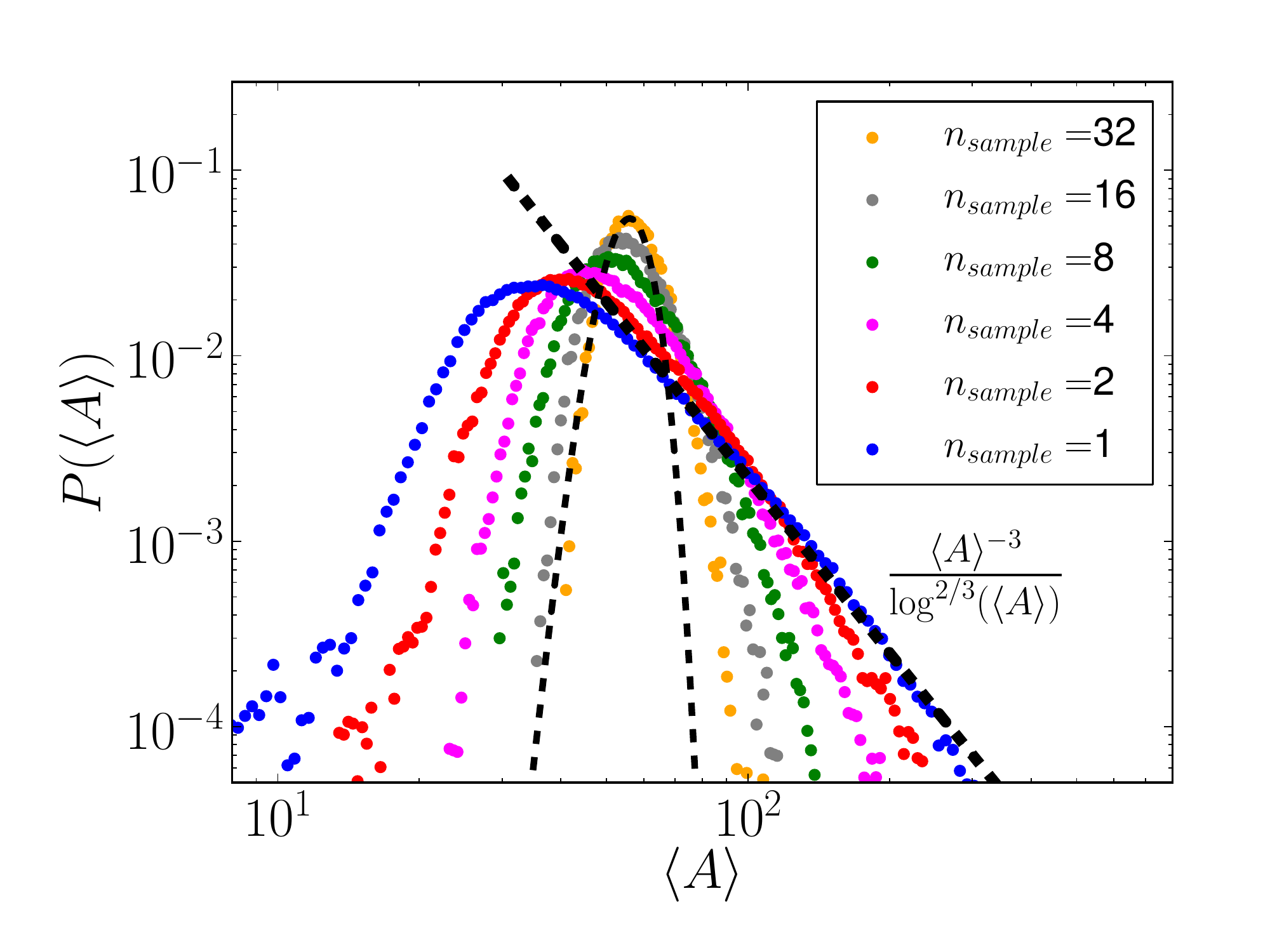}
    \caption{PDF of sample mean $\langle A\rangle$, over $n_{sample}$ realizations (independent ergophages) from the passive nonlinear point-vortex model with parameters $\delta=1$, $\nu/\nu_c=0.15$. For $n_{sample}=1$, the PDF is close to the theoretical prediction for the non-truncated system, and converges to a Gaussian PDF (thin dashed line) as $n_{sample}$ is increased. }
    \label{fig:PDF_conv_to_gauss}
\end{figure}
%%%%%%%%%%%%%%%%%%%%%%%%%%%%%%%%%%%%%%%%%%%%%%

Another prediction of on-off intermittency is that the PDF
of the unstable field shows an \NOTE{integrable} powerlaw divergence at zero \NOTE{amplitude} with an
exponent that approaches the value -1 from above as $\nu\to \nu_c$,
while an exponential cut-off is expected for large values of $A_k$.
%However, even though the time-averaged moments follow the Gaussian scaling, the PDF of each realisation of ergophage amplitudes deviates from the Gaussian noise prediction, as shown in 
Figure \ref{fig:PDF_steady_state} shows the PDF of $A_k$. 
At small values of $A$ the PDF displays a power law $A^\kappa$ 
%range at small $A$ corresponds to the Gaussian on-off prediction 
with $\kappa$ approaching -1 as $\nu\to \nu_c$ in agreement with
the Gaussian on-off prediction. At large $A$ the PDF shows a steeper power-law scaling. % a with exponent close to $-3$. 
% and a logarithmic correction deviates from it. 
In the companion paper \cite{vankan2020extreme}, the asymptotics $P(A)\propto A^{-3} \log^{-2/3}(A)$ at large values of $A$ are derived analytically \NOTE{from a fractional Fokker-Planck equation} \NOTEE{associated with eq. (\ref{eq:A_evolution_inf2})} for non-truncated multiplicative L\'evy noise with parameters $\alpha=2/3$, $\beta=1$ \NOTE{(in the Stratonovich interpretation)}, which fits the present data well. 
%This result differs from the case of additive noise, where the variance is finite, \cite{chechkin2003bifurcation}.
%However, as a consequence of truncation, as shown in figure \ref{fig:PDF_conv_to_gauss} which focusses on this power-law tail far from threshold $\nu/\nu_c=0.15$, averaging $A$ over independent samples leads to a convergence towards a Gaussian distribution.
\NOTE{Since $A^{-1}\log^{-\alpha}(A)$ is only integrable at $A\to \infty$ for $\alpha>1$, the scaling $P(A)\propto A^{-3}\log^{-2/3}(A)$ implies that without a cut-off, only the mean is finite, while the variance and all higher moments diverge.  With a cut-off at length $\epsilon$, all moments are finite, but only the mean is of order one, while all higher moments depend on the cut-off value $\epsilon$, increasing as the latter is decreased. This is an important difference from the Gaussian noise case.}
We note, however, that this difference is diminished as larger samples are used due to the imposed truncation and the law of large numbers. This is demonstrated in figure \ref{fig:PDF_conv_to_gauss} which focuses on this power-law tail far from threshold $\nu/\nu_c=0.15$, and averaging $A$ over independent samples leads to a convergence towards a Gaussian distribution. \NOTE{For a single realization, however, we observe a form close to the theoretical prediction for the non-truncated L\'evy process.}
%
%In our companion paper \cite{vankan2020extreme}, we extend the above analysis of on-off intermittency with L\'evy noise to other values of $\alpha$ and $\tilde{\beta}$ and describe deviations from the critical exponents predicted in the Gaussian case.

\subsection{The fully nonlinear regime}
We now enable ergophages to feed back on the point-vortex flow and include the driving velocity $\mathbf{u}_f$. Initialising a simulation at a \NOTE{condensed} vortex state with $\beta=-1/8$, \NOTEE{$N_v=32$ vortices}, $N_p=32$ ergophages at random locations with small initial amplitudes $A_k$ for given values of $\nu,\delta$ and using a forcing temperature $\beta_f=-1/8$, we let the system evolve in time and measure the mean energy around which the energy fluctuates at late times. \NOTEE{The choice $\beta=\beta_f$ for the initial condition is arbitrary, the system will relax to the same stationary state at late times, independently of what initial condition is chosen. However, since we are interested in the stability of condensate flows, it is a natural initial condition.}

\begin{figure}
    \centering
        \includegraphics[width=8.6cm]{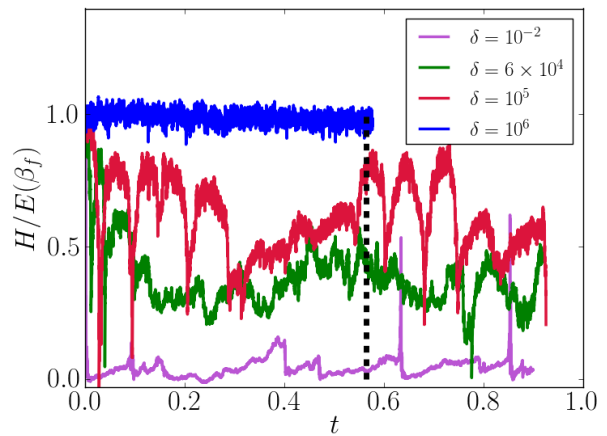}
    \caption{Time series of the 2-D energy $H$\NOTEE{, normalized by the equilibrium energy $E(\beta_f)$ at temperature $\beta_f^{-1}$,} in the fully nonlinear regime at $\nu/\nu_c=0.15$ for different values of $\delta$. At $\delta=10^5$, the flow is close to a 2-D condensate, up to abrupt events when the condensate is disrupted. For decreasing values of $\delta$, ergophages grow to larger amplitudes and lower the energy of the 2-D flow further. \NOTEE{The vertical dashed line indicates the time at which the snapshots in fig. \ref{fig:vis_nl} are taken.}}
    \label{fig:ts_FNL}
\end{figure}

\begin{figure}
    \centering
    \includegraphics[width=8.6cm]{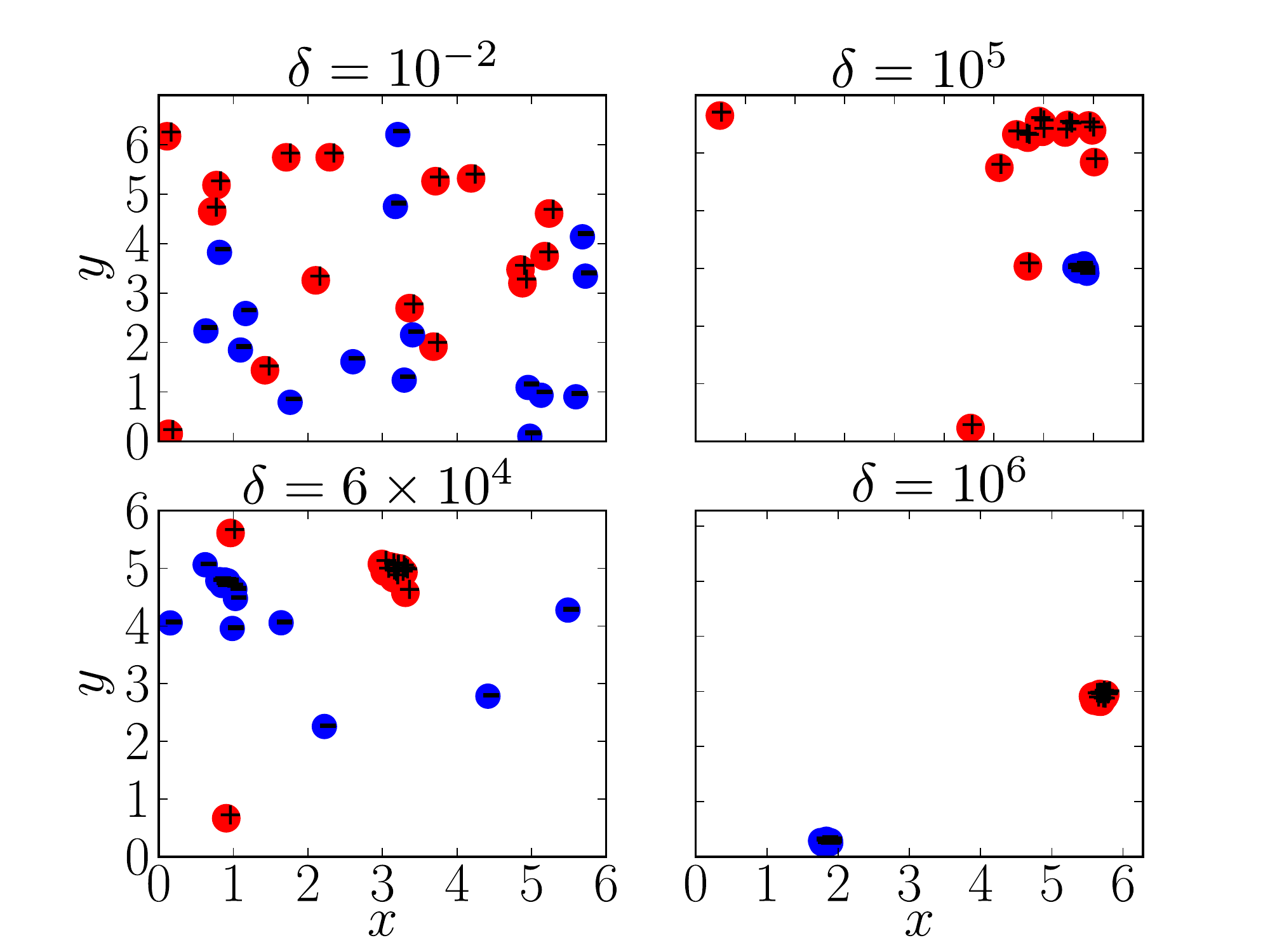}
    \caption{\NOTEE{Snapshots of the point-vortex configuration corresponding to the time indicated by the vertical dashed line in fig. \ref{fig:ts_FNL}. As $\delta$ is decreased, the 3-D perturbations are allowed to grow stronger and disrupt the condensate more and more.}}
    \label{fig:vis_nl}
\end{figure}

Figure \ref{fig:ts_FNL} shows time series of the 2-D energy $H$ in the fully nonlinear regime for $\nu/\nu_c=0.15$ for different values of $\delta$. For large $\delta=10^6$, the 3-D instabilities cannot grow to large amplitudes and therefore do not disrupt the highly energetic condensate. For $\delta=10^5$, a slightly less energetic condensate persists, but is disrupted at random times by catastrophic events which reduce the 2-D flow energy \NOTEE{significantly}, just to rebuild again thanks to the driving. These are the traces of the jumps \NOTEE{ associated with} L\'evy flight dynamics which remain present in the nonlinear regime. Disruptive events occur when an ergophage comes very close to the point-vortex clusters shown in the %first panel of figure \ref{fig:overview_vortices} and the
top panel of figure \ref{fig:visualisaiton_T-8}, \NOTE{extracting the cluster's energy by partially breaking it up}. With decreasing values of $\delta$, the ergophages disrupt the condensate further and further until they reduce its energy to close to zero, driving all point vortices apart. \NOTEE{The snapshots of the point-vortex configurations for different $\delta$ at a fixed time are shown in figure \ref{fig:vis_nl}. They illustrate the gradual disruption of the condensate as $\delta$ is decreased from $\delta=10^6$ to $\delta=10^{-2}$.}

\begin{figure}
    \centering
     \includegraphics[width=8.6cm]{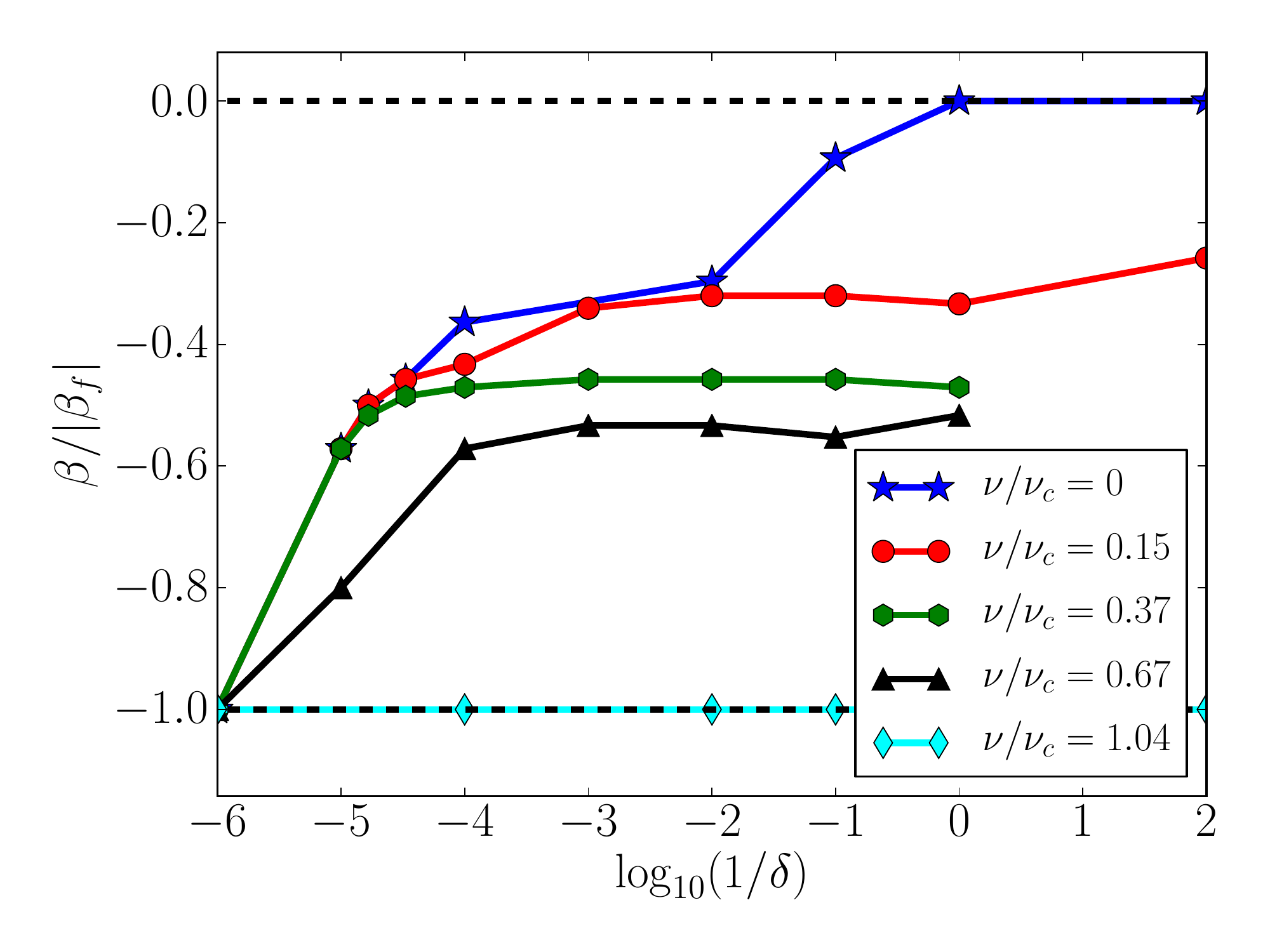} 
    \caption{
    Plot of mean temperature of the point-vortex flow in a fully nonlinear regime in the presence of $N_p=32$ perturbations for varying $\delta$, different curves show different $\nu$. At $\nu/\nu_c>1$, the flow temperature is exactly that of the forcing, i.e. $\beta =\beta_f = -1/8$, since all 3-D perturbations decay. }
    \label{fig:beta_vs_delta_FNL}
\end{figure}

For each simulation, we use the correspondence between mean energy and inverse temperature visualized in figure \ref{fig:E_vs_beta} to assign a vortex temperature \NOTE{based on the measured average point-vortex energy at late times}. We repeat this procedure for several values of $\nu$ and $\delta$ to obtain the diagram shown in figure \ref{fig:beta_vs_delta_FNL}. 
%%%%%%%%%%%%%%%%%%%%%%%%%%%%%%%%%%%%%%%%%%%%%%%%%%%%%%%%%%%
%%%%%%%%%%%%%%%%%%%%%%%%%%%%%%%%%%%%%%%%%%%%%%%%%%%%%%%%%
%%%%%%%%%%%%%%%%%%%%%%%%%%%%%%%%%%%%%%%%%%%%%%%%%%%%%%%%%

%%%%%%%%%%%%%%%%%%%%%%%%%%%%%%%%%%%%%%%%%%%%%%%%%%%%%%%%%%
%
%At $\delta=10^6$, the 2-D condensate is stable, in other words, the 2-D flow is at the forcing temperature, for all values of $\nu$. 
%For $\nu/\nu_c>1$, the 2-D flow is stable and is unaffected by the ergophagues for any value of $\delta$.
For $\nu/\nu_c>1$, 3-D perturbations decay and the 2-D condensate is stable for all values of $\delta$. \NOTEE{As discussed below equation (\ref{eq:uf}), where the driving mechanism is defined, the forced system converges to a finite average energy at late times in the absence of ergophages. In other words, the forcing does not inject a constant energy, but acts rather like a thermostat that aims to maintain the system at a fixed temperature.}
For $\nu/\nu_c<1$, $\beta$ increases with decreasing $\delta$. \NOTEE{This is the onset three-dimensionality, which we characterized in detail in the passive nonlinear regime.}
%as $\delta$ becomes smaller than $10^6$.
%
For smaller values of $\delta$, the perturbation amplitudes saturate at larger values, thus disrupting the 2-D condensate more strongly. When $\delta$ is small enough, the 2-D flow reaches $\beta=0$, which corresponds to a total disruption of the condensate. For $\nu=0$, this occurs at $\delta=1$.  Since the energy-$\beta$ curve shown in figure \ref{fig:E_vs_beta} is very steep at small energies, small deviations in the energy do not necessarily correspond to vanishing $\beta$. Furthermore we note that positive values of $\beta$ induced by the ergophagues were never observed. \NOTEE{Since such states would correspond to flows comprised of long-lived bound vortex-antivortex pairs, the absence thereof is consistent with DNS and experiments of turbulent quasi-2-D flows, where such configurations do not arise spontaneously.}

\NOTEE{The role of the remaining parameters $\beta_f,\epsilon,N_v,N_p$, which do not vary in Fig. \ref{fig:beta_vs_delta_FNL}, is discussed now. Changing $\beta_f$ would alter the 2-D background flow. Decreasing $\beta_f$ would give a more condensed background flow, reducing the surface area of the vortex clusters and thus the chances that an ergophage comes close enough to a cluster to disrupt it. This would require longer simulations and/or larger $N_p$ to obtain reliable statistics. At larger $\beta_f$, the background state ceases to be a condensate, which is undesirable given our focus on condensed base flows. Changing $\epsilon$ would affect the minimum inter-vortex distance in the clusters. Decreasing $\epsilon$, the required time step decreases rapidly according to (\ref{eq:delta_t}), which is numerically challenging, while larger $\epsilon$ would be incompatible with the strong localization of 3-D perturbations. Changing $\epsilon$ also affects the mean growth rate and thus $\nu_c$. Finally, we do not expect $N_v, N_p$ to qualitatively change the system behavior. A larger number of vortices making up the condensate implies more 2-D energy for ergophages to extract. More ergophages, in turn, are more likely to approach the vortex clusters and thus deplete them. Based on the above discussion, while we did not undertake a systematic parameter study, we expect the qualitative model behavior to be robust to parameter changes within appropriate bounds. }

In summary, \NOTEE{above the onset of three-dimensionality, studied in detail in the passive nonlinear case,} the 2-D vortex temperatures depend on the linear and nonlinear damping coefficients of the 3-D flow, ranging from a stable condensate to a complete disruption of the latter. The jump-like L\'evy flight dynamics discussed for of the linear and weakly nonlinear regimes traces through to the nonlinear regime, and shows in the time series in figure \ref{fig:ts_FNL} by a random disruption of the 2-D condensate followed by a rapid subsequent rebuilding of the latter  due to the driving.

%%%%%%%%%%%%%%%%%%%%%%%%%%%%%%%%%%%%%%%%%%%%%%%%%
\section{Conclusions}
\label{sec:conclusions}
%%%%%%%%%%%%%%%%%%%%%%%%%%%%%%%%%%%%%%%%%%%%%%%%%
We have formulated and analyzed a point-vortex model of localized 3-D instabilities on 2-D flows. 
Although the coupling of the 3-D perturbations \NOTEE{to the 2-D flow} in the model is ad-hoc and does not stem directly from the Navier-Stokes equations, it has some 
attractive properties, being energy conserving and reducing to the classical point-vortex model in certain limits. 
%\NOTEE{The spirit of the model is close to shell models of turbulence \cite{biferale2003shell}}. 
Most importantly, the model has led to some very interesting behaviors and predictions that could apply to more realistic quasi-2-D systems exhibiting spectral condensation.

%%% LINEAR
First of all the model predicts fluctuating growth rates with power-law tails, which lead to a L\'evy flight in (logarithmic) perturbation amplitude. \NOTEE{This may be related to recent DNS results \cite{seshasayanan2020onset}, where abrupt, jump-like 3-D instabilities were observed on a strongly condensed, turbulent 2-D background flow. We point out that in \cite{seshasayanan2020onset}, despite the fact that modern GPU computing power was harnessed and after integrating for long times, the time series in their Fig. 1 only contains a few abrupt growth events, far too few to deduce reliable statistical information about the growth rate. This underscores the need for a simplified model like the one presented here, where such information is more readily accessible.} %This suggests a simple explanation for recent DNS results \cite{seshasayanan2020onset} indicating abrupt, jump-like growth of 3-D instabilities on a turbulent 2-D flow. 
Furthermore the model suggests that the onset of the instability depends on the regularization cut-off $\epsilon$. \NOTEE{In realisitic flows, a small-scale cut-off is provided by viscosity.} %For realistic simulations of quasi-two dimensional flows this would imply that the onset of 3-D instabilities would depend on properties measured at the core of the observed vortices that could differ by order of magnitude from mean values. 

A new type of intermittency near the onset of an instability was discovered. The corresponding situation of on-off intermittency in the presence of \NOTEE{ideal, non-truncated L\'evy noise}, is discussed in the companion paper \cite{vankan2020extreme}.

%%% PASSIVE NONLINEAR
In the passive nonlinear regime of the model, we observed a continuous transition from finite to vanishing 3-D amplitudes, with on-off intermittent behavior close to onset. 
%We computed the bifurcation diagram of the moments of the perturbation amplitude from the point-vortex model. %and by integrating a Langevin equation with truncated multiplicative L\'evy noise. 
%As a consequence of the truncation and sample averaging, we recovered the Gaussian scalings near onset. 
However, a deviation from the predictions for Gaussian noise was observed at large values of the \NOTEE{3-D} amplitude, in the form of a power-law tail whose exponent matches theoretical predictions \NOTE{derived from a fractional Fokker-Planck equation in the companion paper \cite{vankan2020extreme}. This exponent also implies that
the saturation amplitude of the second and higher moments would depend on the regularization cut-off $\epsilon$, but not the mean. }
 
%%% FULLY NON-LINEAR

In the fully nonlinear, strongly coupled regime, where the vortex temperature is affected by the presence of perturbations, we characterized the dependence of vortex temperature on the ergophage damping coefficients and showed that at large amplitude of the 3-D perturbations this temperature reduces to zero.
We also showed that at intermediate values of the parameters $\delta$ and $\nu$, a highly energetic condensate, present when 3-D perturbations are small, is disrupted at random times by catastrophic events where 3-D perturbations grow and the condensate amplitude is reduced significantly, after which it recovers. Such events have also been observed in simulations of thin-layer and rotating flows \cite{seshasayanan2020onset,van2019condensates,van2019rare}.  \\

In view of the limitations of existing theories,
%in particular the bounding theories establishing the existence of a threshold height for flow bi-dimensionalisation \cite{gallet2015exact,gallet2015exact2}, 
our model provides a new perspective on 3-D instabilities growing on 2-D flows, which will be useful in analysing and understanding the much more complex results of DNS and potentially guide further theoretical developments.

%%%%%%%%%%%%%%%%%%%%%%%%%%%%%%%%%%%%%%%%%%%%%%%%%
\begin{acknowledgements}
\NOTEE{We thank three anonymous referees for their comments, which helped us improve the clarity of this paper.} We also thank G. Krustolovic for pointing out an important typo in an earlier version of this manuscript. This work was granted access to the HPC resources of MesoPSL financed by the Region 
Ile de France and the project Equip@Meso (reference ANR-10-EQPX-29-01) of the programme Investissements d'Avenir supervised by the Agence Nationale pour la Recherche and the HPC resources of GENCI-TGCC \& GENCI-CINES (Projects No. A0070506421, A0080511423,  A0090506421) where the present numerical simulations have been performed. This work has also been supported by the Agence nationale de la recherche (ANR DYSTURB project No. ANR-17-CE30-0004). AvK acknowledges support by Studienstiftung des deutschen Volkes.
\end{acknowledgements}
%%%%%%%%%%%%%%%%%%%%%%%%%%%%%%%%%%%%%%%%%%5

%%%%%%%%%%%%%%%%%%%%%%%%%%%%%%%%%%%%%%%%%%%%%%%%%
\appendix

%%%%%%%%%%%%%%%%%%%%%%%%%%%%%%%%%%%%%%%%%%%%%%%%%%%%%%%%%%%%%%%%%%
\section{The model equations for periodic boundary conditions} %%%
\label{sec:app_model_in_periodic_bc}                               %%%
%%%%%%%%%%%%%%%%%%%%%%%%%%%%%%%%%%%%%%%%%%%%%%%%%%%%%%%%%%%%%%%%%%
In the main text, the model is presented in infinite space for clarity. Here, we describe the case of 2-D doubly periodic domain $[0,2\pi L]\times [0,2\pi L]$, in which an overall neutral set of an even number $N_v$ of point vortices with circulations $\Gamma_n=(-1)^n\Gamma$, located at positions $\mathbf{x}_v^{(i)}=\left(x_v^{(i)},y_v^{(i)}\right)$ move due to their mutual advection. We describe this configuration using the Weiss-McWilliams formalism introduced in \cite{weiss1991nonergodicity}. In addition, as in the main text, we introduce to $N_p$ localized 3-D perturbations (``ergophages"), idealized as being point-like, at positions $\mathbf{x}_p^{(k)}=\left(x_p^{(k)},y_p^{(k)}\right)$, which are advected by the 2-D point-vortex motions through the 2-D domain, and whose amplitude $A_k$ may grow by extracting energy from the 2-D flow.

%%%%%%%%%%%%%%%%%%%%%%%%%%%%%%%%%%%%%%%%%%%%%%%%%%%%%%%%%%%%%%%%%%
\subsection{Equations of motion and Hamiltonian}
%%%%%%%%%%%%%%%%%%%%%%%%%%%%%%%%%%%%%%%%%%%%%%%%%%%%%%%%%%%%%%%%%%
%The interactions between the two species are taken into account through the velocity field $\mathbf{u}_p(\mathbf{x})$ induced by the 3-D perturbations ("ergophages"), which modifies the motion of the 2-D point vortices. In addition, there may be external driving velocities $\mathbf{u}_f(\mathbf{x})$ and $\mathbf{v}_f$ acting on point vortices and ergophages, respectively. 
The equations of motion of the point vortices and ergophages in the periodic domain are given by the same equations as in the infinite space, (\ref{eq:xv_evolution_inf2}) and (\ref{eq:xp_evolution_inf2}) along with (\ref{eq:def_Uv}).
%\begin{equation}
%    \frac{d}{dt} \mathbf{x}_v^{(i)} =   \mathbf{U}_{v}^{(i)} + \mathbf{U}_{p}^{(i)} + \mathbf{u}_f^{(i)} \label{eq:xv_evolution_perio}
%\end{equation}
%and
%\begin{equation}
%    \frac{d}{dt} \mathbf{x}_p^{(k)} =   \mathbf{U}_{v}^{(k)} + \mathbf{v}_f^{(k)} \label{eq:xp_evolution_perio},
%\end{equation}
%with $\mathbf{U}_p^{(i)}$ being a coupling term, $\mathbf{u}_f^{(i)}$ and $\mathbf{v}_f^{(k)}$ being driving velocities and 
%$$ \mathbf{U}_{v}^{(i)} = \Gamma_i^{-1} \begin{pmatrix}  \partial_{y_v^{(i)}} H \\ - \partial_{x_v^{(i)}} H \end{pmatrix}$$
%the advection by point vortices as before ($\mathbf{U}_v^{(k)}$ is identical, but evaluated at $\mathbf{x}_v^{(i)}\to \mathbf{x}_p^{(k)}$). 
The Hamiltonian in the periodic domain differs from that in the infinite plane, and is given by
\begin{equation}
    H (\lbrace\mathbf{x}_v^{(i)} - \mathbf{x}_v^{(j)}\rbrace)= - \frac{1}{2} \sum_{i,j=1\atop i\neq j}^{N_v}  \Gamma_i \Gamma_j h(\mathbf{x}_v^{(i)} - \mathbf{x}_v^{(j)}), \label{eq:defH_periodic}
\end{equation}
with $\mathbf{x}_{vv}^{ij} \equiv \mathbf{x}_v^{(i)}-\mathbf{x}_v^{(j)} \equiv (x_{vv}^{ij},y_{vv}^{ij})$ and the vortex-pair energy function in the periodic domain given by 
\begin{equation}
    h(x,y) = \sum_{m=-\infty}^\infty \ln \left( \frac{\cosh(x/L-2\pi m)-\cos(y/L)}{\cosh(2\pi m)}\right) - \frac{x^2}{2\pi L^2}, 
    \label{eq:hfunction}
\end{equation}
where the infinite sum over $m$ stems from the sum over all copies of the periodic domain, as shown in \cite{weiss1991nonergodicity}. A useful alternative notation for the 2-D point-vortex advection is given in \cite{weiss1991nonergodicity} as
\begin{equation}
   \Gamma_i^{-1} \begin{pmatrix} + \partial_{y_v^{(i)}} H \\ - \partial_{x_v^{(i)}} H \end{pmatrix} = \sum_{j=1 \atop j\neq i}^{N_v} \Gamma_j \begin{pmatrix} -S\left(y_{vv}^{ij},x_{vv}^{ij}\right) \\ +S\left(x_{vv}^{ij},y_{vv}^{ij}\right) \end{pmatrix},
   \label{eq:alternative_advection_perio}
\end{equation}
in terms of the rapidly converging series 
\begin{equation}
    S(x,y) = \NOTEE{\frac{1}{L}} \sum_{m=-\infty}^\infty \frac{\sin(x/L)}{\cosh(y/L-2\pi m)-\cos(x/L)}.
    \label{eq:defS}
\end{equation}
\NOTE{Equation (\ref{eq:alternative_advection_perio}) relies on the identities $\partial h/\partial x(x,y) = S(x,y)= \partial h/\partial y (y,x)$}.
We note that at small distances, the periodic copies are negligible and one recovers the results valid in the infinite plane. In particular, for $x,y\ll 1$, $S(x,y) \approx xL/(x^2+y^2)$. This enables us to transfer all results pertaining to small distances in the infinite plane to the periodic case.

%--------------------------------------------------------------------
\subsection{Interactions}
%--------------------------------------------------------------------
As in the main text, each of the localized 3-D perturbations is assigned an amplitude $A_k\geq 0$, $k=1,\dots,N_p$, with an associated energy $A_k^2/2$, such that the total energy is again given by (\ref{eq:Etot}), with $H$ given by (\ref{eq:defH_periodic}). For the velocity $\mathbf{U}_p^{(i)}$ induced by the ergophages on the point vortices, we choose again the form given in equation (\ref{eq:Up_tot}). The expression for the dipole field given in equations (\ref{eq:dipole_field}) and (\ref{eq:def_phi}) must be adapted to satisfy the periodic boundary conditions. This is done by tiling $\mathbb{R}^2$ with infinitely many copies of the domain $[0,2\pi L]\times [0,2\pi L]$ and summing over all copies. For a periodic monopole, one obtains 
\begin{equation}
    {\mathbf{u}}_{p,monopole}^{(k)}(\mathbf{x}) = \nabla \phi^k(\mathbf{x}),
\end{equation}
where the potential $\phi^k$, is given by
\begin{equation}
    \phi_k(\mathbf{x}) = h\left(x-x_p^{(k)},y-y_p^{(k)}\right), 
    \label{eq:umonopole_gradphi}
\end{equation}
in terms of the vortex-pair energy function $h(x,y)$ defined in (\ref{eq:hfunction}). %The periodic monopole can also be expressed as 
%\begin{equation}
%    {\mathbf{u}}_{p,monopole}^{(k)}(\mathbf{x}) = \begin{pmatrix} S(x-x_p^{(k)},y-y_p^{(k)}) \\ S(y-y_p^{(k)},x-x_p^{(k)}) \end{pmatrix},
%\end{equation}
%in terms of the rapidly converging series $S(x,y)$ given in equation (\ref{eq:defS}).
The dipole field arises from the difference between two monopoles at small distances, and it is therefore equal to the derivative of the monopole field along the dipole moment $\hat{d}_k = (\cos(\varphi_k),\sin(\varphi_k))$, 
\begin{equation}
    {\mathbf{u}}_p^{(k)}(\mathbf{x}) = (\hat{d}\cdot \nabla_{\mathbf{x}}) \mathbf{u}_{p,monopole}^{(k)}(\mathbf{x})%\begin{pmatrix} S(x-x_p^{(k)},y-y_p^{(k)}) \\ S(y-y_p^{(k)},x-x_p^{(k)}) \end{pmatrix} %\left.\begin{pmatrix} d_k^x \partial_x S(x,y) +d_k^y \partial_x S(x,y) \\ d_k^x \partial_x S(y,x) +d_k^y \partial_y S(y,x) \end{pmatrix} \right|_{(x,y)=\mathbf{\mathbf{x}-\mathbf{x}_p^{(k)}}}. 
    \label{eq:up_dipole_periodic}
\end{equation}

As in the main text, if the $A_k$ obey (\ref{eq:A_evolution_inf2})
%\begin{equation}
%    \frac{\mathrm{d} A_k}{\mathrm{d}t} =  (\gamma_k-\mu) A_k - \delta A^3  \label{eq:A_evolution}
%\end{equation}
%(no implicit summation),
with $\gamma_k$ given by (\ref{eq:defgamma_inf}),
%\begin{equation}
%    \gamma_k = \frac{c'}{2} \sum_{i,j=1\atop i\neq j}^{N_v} \Gamma_i \Gamma_j \nabla h|_{\mathbf{x}_{vv}^{ij}} \cdot\left[ \hat{\mathbf{u}}_p\left(\mathbf{x}_{vp}^{ik}\right)-\hat{\mathbf{u}}_{p}\left(\mathbf{x}_{vp}^{jk}\right)\right],
%    \label{eq:defgamma}
%\end{equation}
then the total energy is conserved in time for arbitrary $\hat{\mathbf{u}}_p$, provided that $\mu=\delta=0$ (no dissipation), and $\mathbf{u}_f=0$. 

The dipole phase $\varphi_k$ is an important degree of freedom, which can be adjusted for sustained growth of ergophage amplitude. Indeed, one can rewrite the growth rate as 
\begin{equation}
    \gamma_k = \Theta_k \cos(\varphi_k) + \Sigma_k \sin(\varphi_k), \label{eq:gamma_k_phase_dependence}
\end{equation}
with 
\begin{align}
\Theta_k =& -\sum_{i=1}^{N_v} \left( \frac{\partial^2 \phi_k(\mathbf{x}_v^{(i)})}{\left(\partial {x_v^{(i)}}\right)^2}  \frac{\partial H}{\partial x_v^{(i)}} + \frac{\partial^2 \phi_k(\mathbf{x}_v^{(i)})}{\partial x_v^{(i)} \partial y_v^{(i)}}  \frac{\partial H}{\partial y_v^{(i)}} \right) 
%\Theta_k=& - \sum_{i=1}^{N_v} \left( \frac{\partial_k S(x_{vp}^{ik},y_{vp}^{ik})}{\partial x_v^{(i)}}  \frac{\partial H}{\partial x_v^{(i)}}  + \frac{\partial S(y_{vp}^{ik},x_{vp}^{ik})}{\partial x_v^{(i)}} \frac{\partial H}{\partial y_v^{(i)}}\right),
\end{align}
and 
\begin{align}
\Sigma_k=& - \sum_{i=1}^{N_v} \left( \frac{\partial^2 \phi_k(\mathbf{x}_v^{(i)})}{\partial x_v^{(i)} \partial y_v^{(i)}}  \frac{\partial H}{\partial x_v^{(i)}}  + \frac{\partial^2 \phi_k(\mathbf{x}_v^{(i)})}{\left(\partial y_
v^{(i)}\right)^2} \frac{\partial H}{\partial y_v^{(i)}}\right).
\end{align}
The form  of (\ref{eq:gamma_k_phase_dependence}) implies that for any vortex configuration, there is an optimum value of the phases $\varphi_k$, for which the growth rate $\gamma_k$ is at its (positive) maximum, is given by  
\begin{equation}
    \varphi_k^{*} = \arctan\left(|\Sigma_k/\Omega_k|\right), \label{eq:opt_phi}
\end{equation}
%The corresponding formulae for the infinite domain read
%\begin{align}
%\begin{pmatrix} \Omega_k \\ \Sigma_k \end{pmatrix} = & c \sum_{i,j=1\atop i\neq j}^{N_v} \Gamma_i \Gamma_j  \left\lbrace \frac{\mathbf{x}_{vv}^{ij}}{|\mathbf{x}_{vv}^{ij}|^2+\epsilon^2}  \left[\frac{1}{|\mathbf{x}_{vp}^{ik}|^2+\epsilon^2}-\frac{1}{|\mathbf{x}_{vp}^{jk}|^2+\epsilon^2}\right]\right. \notag\\ 
%&\hspace{1.8cm} \left.-\frac{2 (\mathbf{x}_{vv}^{ij}\cdot \mathbf{x}_{vp}^{ik})\mathbf{x}_{vp}^{ik}}{(|\mathbf{x}_{vp}^{ik}|^2+\epsilon^2)^2} +\frac{2 (\mathbf{x}_{vv}^{ij}\cdot \mathbf{x}_{vp}^{jk})\mathbf{x}_{vp}^{jk}}{(|\mathbf{x}_{vp}^{jk}|^2+\epsilon^2)^2}\right\rbrace,
%\end{align}
%where $\mathbf{x}_{vp}^{ik}=\mathbf{x}_v^{(i)}-\mathbf{x}_p^{(k)}$. 
The above formulae also apply to dipole ergophages in the infinite domain with the potential (\ref{eq:def_phi}). We let $\varphi_k = \varphi^*_k$ for all $k$ at every instant, implying growth of 3-D instabilities in the inviscid case.

%%%%%%%%%%%%%%%%%%%%%%%%%%%%%%%%%%%%%%%%%%%%%%%%%%%%%%%%%%
\subsection{Numerical implementation of the model}
%%%%%%%%%%%%%%%%%%%%%%%%%%%%%%%%%%%%%%%%%%%%%%%%%%%%%%%%%%
We implemented the equations corresponding to (\ref{eq:xv_evolution_inf2}, \ref{eq:xp_evolution_inf2}, \ref{eq:A_evolution_inf2}) with (\ref{eq:up_dipole_periodic}) and (\ref{eq:opt_phi}) in a fully MPI-parallelized Fortran program using a fourth-order Runge-Kutta time stepper. For the numerical implementation, a regularization was introduced at distances smaller than $\epsilon\ll 2\pi L$, for $\epsilon>0$, in a manner inspired by \cite{krasny1986desingularization}. Specifically, we replace
\begin{equation}
    h(x,y)\to \sum_{m=-\infty}^\infty \ln\left( \frac{\cosh\left(\frac{x-2\pi mL}{L}\right) - \cos\left(\frac{y}{L}\right)+\epsilon^2}{\cosh(2\pi m)} \right) - \frac{x^2}{2\pi L^2}
\end{equation}
and
\begin{equation}
    S(x,y) \to \NOTEE{\frac{1}{L}} \sum_{m=-\infty}^\infty \frac{\sin(x/L)}{\cosh(y/L-2\pi m) -\cos(y/L) + \epsilon^2}.
    \label{eq:regS}
\end{equation}
As mentioned in the main text, the parallelization is implemented straightforwardly by splitting up the sums over vortex-vortex pairs and vortex-parasite pairs into chunks, each of which is assigned to one processor. The choice of the time step is discussed in the main text.
%%%%%%%%%%%%%%%%%%%%%%%%%%%%%%%%%%%%%%%%%%%%%%%%%%%%%%%%%%

\section{Method for generating point-vortex configurations at a given temperature}
\label{sec:appA}
Consider $N$ point vortices located at positions $(x_i,y_i)$, $i=1,\dots,N$ in a given finite domain, with associated Hamiltonian $H$. Pick a positive or negative temperature $T\in \mathbb{R}$. Consider the stochastic gradient dynamics defined by
\begin{align}
    \frac{\mathrm{d}x_i}{\mathrm{d}t} =  - \mathrm{sgn}(T) \frac{\partial H}{\partial x_i} + \sqrt{k_B|T|} \eta^{(1)}_i (t), \label{eq:gradientx} \\
    \frac{\mathrm{d}y_i}{\mathrm{d}t} =  - \mathrm{sgn}(T) \frac{\partial H}{\partial y_i} + \sqrt{k_B|T|} \eta^{(2)}_i(t).  \label{eq:gradienty}
\end{align}
where $\eta_i^{(1)}(t)$ and $\eta_i^{(2)}(t)$ are pairwise independent delta correlated  Gaussian noise terms, i.e. $\langle \eta^{(1)}_i\rangle = \langle \eta^{(2)}_i\rangle  =0$ and $\langle \eta^{(j)}_i(t) \eta^{(j')}_{i'}(t') \rangle = 2 \delta(t-t') \delta_{i,i'} \delta_{j,j'}$, in terms of the ensemble average $\langle \cdot \rangle$. Denote by $\mathbf{X}$ the state vector with entries $X_{2n-1} = x_n$, $X_{2n} = y_n$ for $n=1,\dots,N$. Further, let $\nabla_{\bf X} $ denote the $2N$-dimensional gradient operator with respect to $\mathbf{X}$, then the Fokker-Planck equation for the probability density $P(\mathbf{X},t)$ associated with the given gradient dynamics reads
\begin{equation}
    \partial_t P = \nabla_{\bf X} \cdot \mathbf{F}, \hspace{0.2cm} \text{where } \hspace{0.2cm} \mathbf{F} = \mathrm{sgn}(T) ( \nabla_\mathbf{X}H) P  +  k_B|T|\nabla_\mathbf{X} P. %\equiv \mathrm{sgn}(T) \sum_{i=1}^N \frac{\partial}{\partial x_i} \left[\left(\frac{\partial H}{\partial x_i} P\right)  +\frac{\partial }{\partial y_i} \left( \frac{\partial H}{\partial y_i} P\right)\right] + 2 T \sum_{i=1}^N \left(\frac{\partial^2}{\partial x_i^2} + \frac{\partial^2}{\partial y_i^2}\right) P ,
\end{equation}
In steady state, the flux of probability vanishes if there is no absorption or injection of probability at the boundaries. Solving the zero-flux condition gives the stationary probability density $P_s(\mathbf{X})$ 
\begin{equation}
P_s(\mathbf{X}) = \frac{1}{Z} \exp\left(-\frac{H(\mathbf{X})}{k_BT}\right),    
\end{equation}
which is the Boltzmann equilibrium distribution of the system at temperature $T$. Thus, solving equations (\ref{eq:gradientx}, \ref{eq:gradienty}) numerically, the system reaches a steady state which is precisely the equilibrium at temperature $T$. Importantly, adding the Hamiltonian advection term $U_v^{(i)}$ as in (\ref{eq:xv_evolution_inf2}) does not change this equilibrium, since the associated terms in the Fokker-Planck equation cancel for every index $i$ (being the divergence of a curl).
%%%
\section{Conservaton of energy}
\label{sec:appB}
For the evolution equations (\ref{eq:xv_evolution_inf2}, \ref{eq:A_evolution_inf2}, \ref{eq:defgamma_inf}), for $\mu=\delta=0$ and no forcing, one finds that the total energy is conserved, since 
\begin{align}
    \frac{\mathrm{d} E_{tot}}{\mathrm{d} t} %= &  -   \sum_{i,j = 1\atop i> j}^{N_v} \Gamma_i \Gamma_j \nabla h|_{\mathbf{x}_i-\mathbf{x}_j} \cdot \left\lbrace - \left. \hat{\mathbf{z}} \times \nabla \psi'\right|_{\mathbf{x_i}} + \left. \hat{\mathbf{z}} \times \nabla \psi'\right|_{\mathbf{x_j}} + \sum_{r=1}^{N_p} \left[\mathbf{u}_r(\mathbf{x}_i) -\mathbf{u}_r(\mathbf{x}_j) \right]\right\rbrace + \sum_{k=1}^{N_p} A_k \frac{\mathrm{d} A_k}{\mathrm{d} t}\\
    =&  \frac{dH}{dt} + \sum_{k=1}^{N_p} A_k \frac{dA_k}{dt} \\%  c\sum_{k=1}^{N_p} \sum_{i,j = 1\atop i> j}^{N_v}  \Gamma_i \Gamma_j \nabla h|_{\mathbf{x}_{vv}^{ij}} \cdot \left( \hat{\mathbf{u}}_p\left(\mathbf{x}_{vp}^{ik}\right)  - \hat{\mathbf{u}}_p\left(\mathbf{x}_{vp}^{jk}\right) \right) A_k^2 \notag \\ + &\sum_{k=1}^{N_p} A_k \frac{\mathrm{d} A_k}{\mathrm{d} t}  \\
    =& \sum_{i=1}^{N_v} \mathbf{U}_p^{(i)} \cdot \nabla_{\mathbf{x}_v^{(i)}} H + \sum_{k=1}^{N_p} A_k (\gamma_k A_k) \\%-   c\sum_{k=1}^{N_p} \sum_{i,j = 1\atop i> j}^{N_v}  \Gamma_i \Gamma_j \nabla h|_{\mathbf{x}_{vv}^{ij}} \cdot \left( \hat{\mathbf{u}}_p\left(\mathbf{x}_{vp}^{ik}\right)  - \hat{\mathbf{u}}_p\left(\mathbf{x}_{vp}^{jk}\right) \right) A_k^2 \notag \\
    =& \sum_{i=1}^{N_v} \sum_{k=1}^{N_p} A_k^2 \mathbf{u}_p^{(k)}(\mathbf{x}_v^{(i)})\cdot \nabla_{\mathbf{x}_v^{(i)}} H \notag \\ &- \sum_{k=1}^{N_p}\sum_{i=1}^{N_v} A_k^2 \mathbf{u}_p^{(k)}(\mathbf{x}_v^{(i)})\cdot \nabla_{\mathbf{x}_v^{(i)}} H\\
    =& 0% c\sum_{k=1}^{N_p} \sum_{i,j = 1\atop i> j}^{N_v}  \Gamma_i \Gamma_j \nabla h|_{\mathbf{x}_{vv}^{ij}} \cdot \left( \hat{\mathbf{u}}_p\left(\mathbf{x}_{vp}^{ik}\right)  - \hat{\mathbf{u}}_p\left(\mathbf{x}_{vp}^{jk}\right) \right) A_k^2
\end{align}
This conservation of energy is independent of the modelling choice of the velocity field ${\mathbf{u}}_p$ and of the particular form of the Hamiltonian. Hence the conservation holds for arbitrary boundary conditions.

%------------------------------------------------------------------------------
\section{Vanishing mean growth rate for monopole 3-D perturbations and derivation of dipole formulas}
\label{sec:appC}
\begin{figure}
    \centering
    \includegraphics[width=8.6cm]{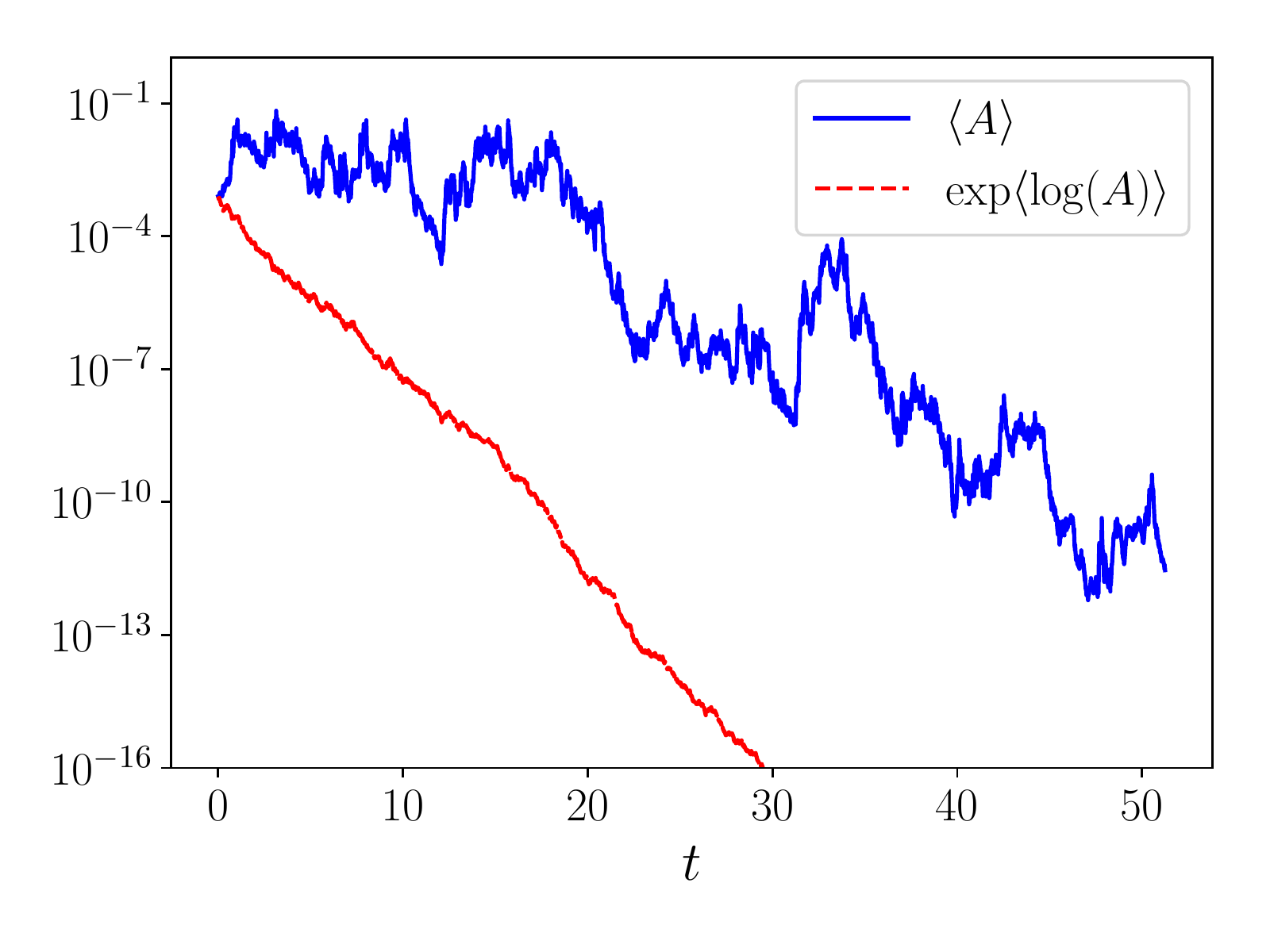}
    \caption{Lin-log plot of the time series of the first moment $M_1=\langle A \rangle$ and the zeroth moment $M_0=\exp(\langle \log(A)\rangle)$ of ergophage amplitude, in terms of the sample average $\langle f(A)\rangle= \frac{1}{N_p} \sum_k f(A_k)$, from a passive nonlinear simulation with $N_p=32$ ergophages inducing a monopole field, experiencing disspation $\nu,\delta>0$. The zeroth moment decays exponentially, indicating that the mean growth rate is negative. Both moments clearly decay at late times as predicted theoretically.}
    \label{fig:monopole_amp_decay}
\end{figure}

The simplest possible choice for the velocity induced by 3-D perturbations, $\mathbf{u}_p(\mathbf{x})$, in infinite space is an isotropic radial profile,
\begin{equation}
    {\mathbf{u}}_p^{(k)}(\mathbf{x}) = \frac{\mathbf{x}-\mathbf{x}_p^{(k)}}{|\mathbf{x}-\mathbf{x}_p^{(k)}|^2},
\end{equation}
i.e. a monopole profile. Since it decays at infinity, it is admissible in the infinite plane. In a periodic domain, however, it needs to be adapted to the boundary conditions by summing over an infinite grid of images: 
\begin{align}
    {\mathbf{u}}_p(\mathbf{x})^{(k)} =& \sum\limits_{n,m=-\infty}^\infty \frac{\mathbf{x}-\mathbf{x}_p^{(k)}-\begin{pmatrix}2\pi n \\2\pi m\end{pmatrix}}{\left|\mathbf{x}-\mathbf{x}_p^{(k)}-\begin{pmatrix}2\pi nL \\2\pi mL \end{pmatrix}\right|^2} \notag \\
    =& \begin{pmatrix} S(x-x_p^{(k)},y-y_p^{(k)}) \\ S(y-y_p^{(k)},x-x_p^{(k)}) \end{pmatrix},
    \label{eq:upmonopole_alternative}
\end{align}
where $S(x,y)$ is as defined by the rapidly converging series given in (\ref{eq:defS}) and regularized in (\ref{eq:regS}). Equation (\ref{eq:upmonopole_alternative}) provides an alternative expression for the periodic monopole field, equivalent to that in (\ref{eq:umonopole_gradphi}). We note that the infinite sum is exactly the double series calculated by Weiss and McWilliams in \cite{weiss1991nonergodicity}. The corresponding growth rate of perturbation $k$ given in (\ref{eq:defgamma_inf}) can be rewritten as
\begin{align}
    \gamma_k =& %\frac{c}{2} \sum\limits_{i,j=1 \atop i\neq j}^{N_p} \Gamma_i \Gamma_j \nabla h|_{\mathbf{x}_{vv}^{ij}} \cdot \left[ \hat{\mathbf{u}}\left(\mathbf{x}_{vp}^{ik}\right)-\hat{\mathbf{u}}\left(\mathbf{x}_{vp}^{jk}\right) \right] \notag \\
     \frac{c}{2} \sum\limits_{i,j=1 \atop i\neq j}^{N_p} \Gamma_i \Gamma_j \nabla h|_{\mathbf{x}_{vv}^{ij}} \cdot \left(\left.\nabla h\right|_{\mathbf{x}_{vp}^{ik}} -  \left. \nabla h\right|_{\mathbf{x}_{vp}^{jk}} \right) %\left[ \begin{pmatrix} S(x_{vp}^{ik},y_{vp}^{ik})-S(x_{vp}^{jk},y_{vp}^{jk}) \\ S(y_{vp}^{ik},x_{vp}^{ik})-S(y_{vp}^{jk},x_{vp}^{jk})\end{pmatrix}\right],
     \label{eq:gamma_monopole}
\end{align}
\NOTE{with $\mathbf{x}_{vv}^{ij} = \mathbf{x}_v^{(i)}-\mathbf{x}_{v}^{(j)}$ and $\mathbf{x}_{vp}^{ik} = \mathbf{x}_v^{(i)}-\mathbf{x}_{p}^{(k)} $. It has been used that  from eq. (\ref{eq:defS}) that $\partial h /\partial x (x,y) =S(x,y)= \partial h /\partial y (y,x)$. For simplicity, since the sum is over vortex pairs, consider a single such pair with circulations $\Gamma_1,\Gamma_2$ at arbitrary positions $\mathbf{x}_1,\mathbf{x}_2$.} %(by symmetry the orientation of the pair should not matter). 
Place a single ergophage at position $(x,y)$. The sum over $i,j$ in (\ref{eq:gamma_monopole}) reduces to a single term. 
%\begin{align}
%    \gamma = \frac{c \Gamma_1\Gamma_2}{2} f(d) \left[\sum_{m=-\infty}^\infty \frac{\sin(\frac{d-y}{L})}{\cosh(\frac{-x+2\pi mL}{L}) - \cos(\frac{d-y}{L})+\epsilon^2} \right. \notag \\ \left. + \frac{\sin(\frac{d+y}{L})}{\cosh(\frac{-x+2\pi mL}{L}) - \cos(\frac{d+y}{L})+\epsilon^2} \right], 
%\end{align}
%where $f(d) = \sum\limits_{n=-\infty}^{\infty} \frac{\sin(2d/L)}{\cosh(2\pi n)-\cos(2d/L)} $ is independent of the ergophage position. 
Applying the averaging operator over ergophage positions,
$$\overline{F}\equiv \frac{1}{4\pi^2 L^2}\int_0^{2\pi L} \int_0^{2\pi L }   F(x,y) dx dy,  $$
to the growth rate gives zero, since $h$ is $2\pi L$-periodic in both the $x$ and $y$ directions.
%\begin{align}
%    \overline{\gamma} =& \frac{c\Gamma_1\Gamma_2}{8\pi^2}f(d) \left[ \int_0^{2\pi}  \int_0^{2\pi}  [\sin(d-y) g(x,d-y) \right.  \notag \\ & \hspace{2.5cm} \left. \phantom{\int_0^2} + \sin(d+y) g(x,d+y)] dx dy \right],
%\end{align}
%where $g(x,y)=\sum\limits_{m=-\infty}^\infty[\cosh(-x+2\pi m)-\cos(y)+\epsilon^2]^{-1}$ is an even function of $y$. A change of variables yields
%\begin{align}
%    \overline{\gamma} \propto &  \int_0^{2\pi} dx \int_{0}^{2\pi}  dy [-\sin(y) g(x,y) + \sin(y) g(x,y)] \notag \\  &= 0.
%\end{align}
We conclude that the mean growth rate of a monopole ergophage due to a single vortex pair vanishes, for arbitrary vortex positions. Thus the mean total ergophage growth rate, being the sum of pair contributions, also vanishes. Assuming that for a given vortex configuration, all ergophage positions are equally likely, then the resulting mean growth rate vanishes in the absence of dissipation. When dissipation is added, then 3-D perturbations must decay at long times. This is illustrated by a long run with $N_p=32$ passive nonlinear monopole ergophages and $N_v=32$ point vortices in figure \ref{fig:monopole_amp_decay}. Therefore, the monopole model is insufficient and the dipole model suggests itself as having the minimal complexity to capture mean growth of 3-D perturbations. %The formula for the dipole can be obtained by considering the leading order difference between two monopole perturbations at a small distance $r$ along a direction $\hat{d}=(\cos(\phi),\sin(\phi))$, which, divided by $r$, amounts to the derivative of $S(x,y)$ in the direction of $\hat{d}$. This directly leads to the formula given in the equation (\ref{eq:up_dipole_periodic}) when the explicit expression (\ref{eq:defS}) is substituted for $S(x,y)$.

%%%%%%%%%%%%%%%%%%%%%%%%%%%%%%%%%%%%%%%%%%%%%%%%%%%%%%%%%%%
\section{Power laws in growth rate probability density}  %%
\label{sec:appD}                                         %%
%%%%%%%%%%%%%%%%%%%%%%%%%%%%%%%%%%%%%%%%%%%%%%%%%%%%%%%%%%%
For the dipole parasites introduced in the main text, consider the growth rate of the amplitude of a given ergophage at location $\mathbf{x}_p$, associated with a vortex pair of circulation $\Gamma_1,\Gamma_2$ at positions $\mathbf{x}_1=(\ell/2,0),\mathbf{x}_2=(-\ell/2,0)$. %\CMNT{$\ell$ is a bad choise of symbol, its to close to the dipole moment $\hat{d}$ and make the prodact $dr=d*r$ look like an infinitesimal increment of $r$. }. 
We are interested in the tails of the probability density function (PDF), where the ergophage is very close to one or several point vortices, hence boundary conditions are irrelevant and we perform the analysis in the infinite plane. The localized perturbation has a dipole moment $\hat{d}=(\cos(\varphi),\sin(\varphi))$ attached to it as well as an amplitude $A$, whose growth rate is given by
\begin{align}
 \gamma =  \frac{\Gamma_1 \Gamma_2}{\ell} &\left[ \frac{\cos(\varphi)}{(x-\frac{\ell}{2})^2+y^2} \right. \notag \\
 &- \frac{2[(\frac{\ell}{2}-x)\cos(\varphi)-y\sin(\varphi)](\frac{\ell}{2}-x)}{[(x-\frac{\ell}{2})^2+y^2]^2} \notag
\\&- \frac{\cos(\varphi)}{(x+\frac{\ell}{2})^2+y^2}  \notag \\ &+\left.\frac{2[(\frac{\ell}{2}+x)\cos(\varphi)+y\sin(\varphi)](\frac{\ell}{2}+x)}{[(\frac{\ell}{2}+x)^2+y^2]^2}\right].  
\end{align}
There are two limits of interest to be considered, namely the \textit{dilute} limit corresponding to small inverse vortex temperatures $|\beta|\ll 1$ and the \textit{dense} limit corresponding to large(-magnitude) inverse vortex temperatures, i.e. pairs of opposite-sign vortices for $\beta>0$ and clusters of same-sign vortices for $\beta<0$.
%%%%%%%%%%%%%%%%%%%%%%%%%%%%%%%%%%
\subsection{The dilute limit}  %%%
%%%%%%%%%%%%%%%%%%%%%%%%%%%%%%%%%%
In this case, the tails of the PDF of $\gamma$ are generated by events in which the perturbation is closer to a single point vortex than to any other vortices, i.e. $\mathbf{x}_p = \mathbf{x}_1 + r (\cos(\phi),\sin(\phi))$, $r\ll \ell$.  In this case, 
\begin{align}
\gamma &\sim \frac{\Gamma_1 \Gamma_2}{\ell r^2} \left[ \sin(\varphi)\sin(2\theta) - \cos(\varphi) \cos(2\theta) \right] \notag\\&= -\frac{\Gamma_1\Gamma_2}{\ell r^2} \cos(2\theta+\varphi).
\end{align}
Since we consider the case where $\varphi$ is optimal at every position, one finds $\varphi =-2\theta + n \pi$, $n\in \mathbb{N}$ and
\begin{equation}
\gamma \sim \frac{|\Gamma_1\Gamma_2|}{\ell r^2} \hspace{ 0.5cm} \Leftrightarrow \hspace{0.5cm} r(\gamma) \sim \sqrt{\frac{\gamma \ell}{|\Gamma_1 \Gamma_2|}} \label{dilute_gamma}
\end{equation}
Assuming that all ergophage positions are equally probable, 
\NOTE{then the probability of of being at distance between $r$ and $r+dr$ is proportional to the ring area $2\pi r dr$. This can be inverted using (\ref{dilute_gamma}) to obtain} a prediction for the PDF of $\gamma$, namely 
\begin{equation}
    P(\gamma)  = r(\gamma) \frac{dr(\gamma)}{d\gamma}\propto \frac{1}{\gamma^2}
\end{equation} 
%%%%%%%%%%%%%%%%%%%%%%%%%%%%%%%%
\subsection{The dense limit}  %%
%%%%%%%%%%%%%%%%%%%%%%%%%%%%%%%%
In this case, the tails of the PDF of the growth rate stem from encounters of the localized perturbation with pairs of vortices, i.e. $\mathbf{x}_p = r(\cos(\theta),\sin(\theta))$, $r\gg \ell$. Then, one finds at leading order in $\ell$ that
\begin{align*}
\gamma \sim&  \frac{\Gamma_1 \Gamma_2}{\ell} \left( -2\ell\frac{\cos(\varphi)\cos(\theta)}{r^3} \right. \\ 
&\left.   \hspace{1cm}+  2\ell  \frac{ y \sin(\varphi) (y^2-3x^2) - 2x \cos(\varphi) (x^2-x^2)}{r^6} \right) 
 \\ &  \sim -\frac{ 2\Gamma_1 \Gamma_2 }{r^3}  \cos(3\theta -\varphi)
\end{align*}
Again assuming that $\varphi$ is optimal, then $\varphi = -3\theta +n \pi $, $n\in \mathbb{N}$, such that
$$ \gamma \sim \frac{2 |\Gamma_1 \Gamma_2|}{r^3},$$
which leads to the growth rate PDF, again under the assumption that all ergophage positions are equally probable
$$P(\gamma) =r(\gamma) \frac{dr(\gamma)}{d\gamma}\propto \frac{1}{\gamma^{\frac{1}{3}+\frac{4}{3}}} = \frac{1}{\gamma^{5/3}}, $$
with an exponent $-5/3$, whose magnitude is less than $2$. For both cases (dense and dilute), the PDF has neither a finite mean, nor a finite variance. We note that the exponent $-5/3$ found here bears no relation to Kolmogorov's spectral exponent, it is merely a consequence of the modelling choices made.

\bibliography{biblio}
%\bibliography{report}
\end{document}